\documentclass[journal,10pt,twocolumn,final]{IEEEtran}
\usepackage{graphicx}
\usepackage{amsmath,amssymb,amsthm,graphicx,psfrag,url,color}
\usepackage[caption=false,font=footnotesize]{subfig}
\usepackage{comment}
\usepackage{mathdots}

\newcommand{\bc}{\begin{center}}
\newcommand{\ec}{\end{center}}

\newcommand{\ba}{\begin{array}}
\newcommand{\ea}{\end{array}}

\newcommand{\bt}{\begin{tabular}}
\newcommand{\et}{\end{tabular}}

\newcommand{\eems}{\end{multline*}}
\newcommand{\eem}{\end{multline}}
\newcommand{\eeams}{\end{eqnarray}\vspace{0.1cm}}
\newcommand{\eearms}{\end{eqnarray*}\vspace{0.1cm}}
\newcommand{\edmms}{\end{displaymath}\vspace{0.1cm}}

\newcommand{\bems}{\begin{multline*}}
\newcommand{\bem}{\begin{multline}}
\newcommand{\beams}{\vspace{0.1cm} \begin{eqnarray}}
\newcommand{\bearms}{\vspace{0.1cm} \begin{eqnarray*}}
\newcommand{\bdmms}{\vspace{0.1cm} \begin{displaymath}}
\newcommand{\E}{\mathrm{E}}
\newcommand{\Var}{\mathrm{Var}}
\newcommand{\be}{\begin{equation}}
\newcommand{\ee}{\end{equation}}

\newcommand{\bes}{\begin{equation*}}
\newcommand{\ees}{\end{equation*}}

\newcommand{\bea}{\begin{eqnarray}}
\newcommand{\eea}{\end{eqnarray}}

\newcommand{\bear}{\begin{eqnarray*}}
\newcommand{\eear}{\end{eqnarray*}}

\newcommand{\bdm}{\begin{displaymath}}
\newcommand{\edm}{\end{displaymath}}

\newcommand{\bit}{\begin{itemize}}
\newcommand{\eit}{\end{itemize}}

\newcommand{\ben}{\begin{enumerate}}
\newcommand{\een}{\end{enumerate}}

\newcommand{\bd}{\begin{document}}
\newcommand{\ed}{\end{document}}

\newcommand{\psp}{\hspace{0.6cm}}
\newcommand{\hspone}{\hspace{1cm}}
\newcommand{\hsptwo}{\hspace{2cm}}

\newcommand{\hspthree}{\hspace{3cm}}
\newcommand{\hspfour}{\hspace{4cm}}

\newcommand{\hspfive}{\hspace{5cm}}
\newcommand{\hspsix}{\hspace{6cm}}

\newcommand{\hspseven}{\hspace{7cm}}
\newcommand{\hspeight}{\hspace{8cm}}

\newcommand{\vspone}{\vspace{1cm}}
\newcommand{\vsptwo}{\vspace{2cm}}

\newcommand{\vspthree}{\vspace{3cm}}
\newcommand{\vspfour}{\vspace{4cm}}

\newcommand{\vspfive}{\vspace{5cm}}
\newcommand{\vspsix}{\vspace{6cm}}

\newcommand{\vspseven}{\vspace{7cm}}
\newcommand{\vspeight}{\vspace{8cm}}

\newcommand{\vspnine}{\vspace{9cm}}
\newcommand{\vspten}{\vspace{10cm}}

\newcommand{\vspeleven}{\vspace{11cm}}
\newcommand{\vsptwelve}{\vspace{12cm}}

\newcommand{\qett}{\mbox{$(q^{-1})$}}
\newcommand{\zett}{\mbox{$(z^{-1})$}}

\newcommand{\qo}{q^{-1}}
\newcommand{\zettu}{\mbox{$z^{-1}$}}

\newcommand{\linf}[1]{{#1}\rightarrow \infty}

\newcommand{\cf}[1]{(\ref{#1})}

\newcommand{\eiw} {\mathrm{e}^{\mathrm{i}\omega}}
\newcommand{\emiw}{\mathrm{e}^{-\mathrm{i}\omega}}

\newcommand{\diag}{\mbox{${\rm diag}$}}
\newcommand{\cov}{\mbox{${\rm Cov}$}}
\newcommand{\Ex}{{\sf E}}
\newcommand{\erm}{\mathrm{e}}
\newcommand{\irm}{\mathrm{i}}
\newcommand{\Prj} {{\bf \Pi}^\bot}
\newcommand{\nn}{$\vert\mathcal{N}\vert$}
\newcommand{\enn}{$\vert\hat{\mathcal{N}}\vert$}
\newcommand{\bA} {{\bf A}}
\newcommand{\bB} {{\bf B}}
\newcommand{\bC} {{\bf C}}
\newcommand{\bD} {{\bf D}}
\newcommand{\bE} {{\bf E}}
\newcommand{\bF} {{\bf F}}
\newcommand{\bG} {{\bf G}}
\newcommand{\bH} {{\bf H}}
\newcommand{\bI} {{\bf I}}
\newcommand{\bJ} {{\bf J}}
\newcommand{\bK} {{\bf K}}
\newcommand{\bL} {{\bf L}}
\newcommand{\bM} {{\bf M}}
\newcommand{\bN} {{\bf N}}
\newcommand{\bO} {{\bf O}}
\newcommand{\bP} {{\bf P}}
\newcommand{\bQ} {{\bf Q}}
\newcommand{\bR} {{\bf R}}
\newcommand{\bS} {{\bf S}}
\newcommand{\bT} {{\bf T}}
\newcommand{\bU} {{\bf U}}
\newcommand{\bV} {{\bf V}}
\newcommand{\bW} {{\bf W}}
\newcommand{\bY} {{\bf Y}}
\newcommand{\bX} {{\bf X}}
\newcommand{\bZ} {{\bf Z}}

\newcommand{\GLm} {{\bf \Lambda}}
\newcommand{\GSg} {{\bf \Sigma}}
\newcommand{\GPh} {{\bf \Phi}}
\newcommand{\GPs} {{\bf \Psi}}
\newcommand{\GPi} {{\bf \Pi}}
\newcommand{\GDl} {{\bf \Delta}}
\newcommand{\GGm} {{\bf \Gamma}}
\newcommand{\GUp} {{\bf \Upsilon}}
\newcommand{\GOm} {{\bf \Omega}}
\newcommand{\GTh} {{\bf \Theta}}
\newcommand{\GXi} {{\bf \Xi}}

\newcommand{\gal} {\boldsymbol{\alpha}}
\newcommand{\gbt} {\boldsymbol{\beta}}
\newcommand{\ggm} {\boldsymbol{\gamma}}
\newcommand{\gdl} {\boldsymbol{\delta}}
\newcommand{\gep} {\boldsymbol{\epsilon}}
\newcommand{\vep} {\boldsymbol{\varepsilon}}
\newcommand{\gzt} {\boldsymbol{\zeta}}
\newcommand{\get} {\boldsymbol{\eta}}
\newcommand{\gth} {\boldsymbol{\theta}}
\newcommand{\vth} {\boldsymbol{\vartheta}}
\newcommand{\git} {\boldsymbol{\iota}}
\newcommand{\gkp} {\boldsymbol{\kappa}}
\newcommand{\glm} {\boldsymbol{\lambda}}
\newcommand{\gmu} {\boldsymbol{\mu}}
\newcommand{\gnu} {\boldsymbol{\nu}}
\newcommand{\gxi} {\boldsymbol{\xi}}
\newcommand{\gpi} {\boldsymbol{\pi}}
\newcommand{\vpi} {\boldsymbol{\varpi}}
\newcommand{\grh} {\boldsymbol{\rho}}
\newcommand{\vrh} {\boldsymbol{\varrho}}
\newcommand{\gsg} {\boldsymbol{\sigma}}
\newcommand{\vsg} {\boldsymbol{\varsigma}}
\newcommand{\gtu} {\boldsymbol{\tau}}
\newcommand{\gup} {\boldsymbol{\upsilon}}
\newcommand{\gph} {\boldsymbol{\phi}}
\newcommand{\vph} {\boldsymbol{\varphi}}
\newcommand{\gch} {\boldsymbol{\chi}}
\newcommand{\gps} {\boldsymbol{\psi}}
\newcommand{\gom} {\boldsymbol{\omega}}

\newcommand{\Ba}  {{\bf a}}
\newcommand{\Bb}  {{\bf b}}
\newcommand{\Bc}  {{\bf c}}
\newcommand{\Bd}  {{\bf d}}
\newcommand{\Be}  {{\bf e}}
\newcommand{\Bf}  {{\bf f}}
\newcommand{\Bg}  {{\bf g}}
\newcommand{\Bh}  {{\bf h}}
\newcommand{\Bi}  {{\bf i}}
\newcommand{\Bj}  {{\bf j}}
\newcommand{\Bk}  {{\bf k}}
\newcommand{\Bl}  {{\bf l}}
\newcommand{\Bm}  {{\bf m}}
\newcommand{\Bn}  {{\bf n}}
\newcommand{\Bo}  {{\bf o}}
\newcommand{\Bp}  {{\bf p}}
\newcommand{\Bq}  {{\bf q}}
\newcommand{\Br}  {{\bf r}}
\newcommand{\Bs}  {{\bf s}}
\newcommand{\Bt}  {{\bf t}}
\newcommand{\Bu}  {{\bf u}}
\newcommand{\Bv}  {{\bf v}}
\newcommand{\Bw}  {{\bf w}}
\newcommand{\Bx}  {{\bf x}}
\newcommand{\By}  {{\bf y}}
\newcommand{\Bz}  {{\bf z}}

\newcommand{\za} {\boldsymbol{a}}
\newcommand{\zb} {\boldsymbol{b}}
\newcommand{\zc} {\boldsymbol{c}}
\newcommand{\zd} {\boldsymbol{d}}
\newcommand{\ze} {\boldsymbol{e}}
\newcommand{\zf} {\boldsymbol{f}}
\newcommand{\zg} {\boldsymbol{g}}
\newcommand{\zh} {\boldsymbol{h}}
\newcommand{\zi} {\boldsymbol{i}}
\newcommand{\zj} {\boldsymbol{j}}
\newcommand{\zk} {\boldsymbol{k}}
\newcommand{\zl} {\boldsymbol{\ell}}
\newcommand{\zm} {\boldsymbol{m}}
\newcommand{\zn} {\boldsymbol{n}}
\newcommand{\zo} {\boldsymbol{o}}
\newcommand{\zp} {\boldsymbol{p}}
\newcommand{\zq} {\boldsymbol{q}}
\newcommand{\zr} {\boldsymbol{r}}
\newcommand{\zs} {\boldsymbol{s}}
\newcommand{\zt} {\boldsymbol{t}}
\newcommand{\zu} {\boldsymbol{u}}
\newcommand{\zv} {\boldsymbol{v}}
\newcommand{\zw} {\boldsymbol{w}}
\newcommand{\zx} {\boldsymbol{x}}
\newcommand{\zy} {\boldsymbol{y}}
\newcommand{\zz} {\boldsymbol{z}}

\newcommand{\hide}[1] {}

\newtheorem{Thm}{Theorem}
\newtheorem{Pro}{Proposition}
\newtheorem{Defn}{Definition}
\newtheorem{Exm}{Example}
\newtheorem{Lem}{Lemma}
\newtheorem{Asm}{Assumption}
\newtheorem{Cor}{Corollary}
\newtheorem{Alg}{Algorithm}
\newtheorem{paRem}{Remark}
\newenvironment{Rem}{\begin{paRem}\rm }{\end{paRem}}

\newcommand{\ColvecTwo}[2]{\left[\ba{c} {#1} \\ {#2} \ea \right]}
\newcommand{\colvecTwo}[2]{\left[\ba{cc} {#1} & {#2} \ea \right]^\top}

\newcommand{\colvecThr}[2]{\left[\ba{ccc} {#1} & \cdots & {#2} \ea \right]^\top}
\newcommand{\ColvecThr}[2]{\left[\ba{c} {#1} \\ \vdots \\ {#2} \ea \right]}
\newcommand{\rowvecThr}[2]{\left[\ba{ccc} {#1} & \cdots & {#2} \ea \right]}
\newcommand{\MatThrTwo}[4]{\left[\ba{ccc} {#1}  & {#3} \\ \vdots & \vdots \\ {#2} & {#4} \ea \right]}

\begin{document}
\author{Shah Mahdi Hasan, Kaushik Mahata and Md Mashud Hyder}
\title{Fast Uplink Grant-Free NOMA with Sinusoidal Spreading Sequences}
\maketitle
\begin{abstract}
Uplink (UL) dominated sporadic transmission and stringent latency requirement of massive machine type communication (mMTC) forces researchers to abandon complicated grant-acknowledgment based legacy networks. UL grant-free non-orthogonal multiple access (NOMA) provides an array of features which can be harnessed to efficiently solve the problem of massive random connectivity and latency. Because of the inherent sparsity in user activity pattern in mMTC, the trend of existing literature specifically revolves around compressive sensing based multi user detection (CS-MUD) and Bayesian framework paradigm which employs either random or Zadoff-Chu spreading sequences for non-orthogonal multiple access. In this work, we propose sinusoidal code as candidate spreading sequences. We show that, sinusoidal codes allow some non-iterative  algorithms to be employed in context of active user detection, channel estimation and data detection in a UL grant-free mMTC system. This relaxes the requirement of several impractical assumptions considered in the state-of-art algorithms with added advantages of performance guarantees and lower computational cost. Extensive simulation results validate the performance potential of sinusoidal codes in realistic mMTC environments.  
\end{abstract}

\begin{IEEEkeywords}
massive machine-type communication (mMTC), grant-free, active user detection, channel estimation, non-orthogonal multiple access (NOMA), subspace estimation
\end{IEEEkeywords}

\section{Introduction}
\label{intro}
\subsection{Background and motivation}


Under the IMT 2020 Vision Framework of International Telecommunication Union (ITU) \cite{metis}
the massive machine-type communication (mMTC) is a major service category of a 5G system. A typical mMTC network consists of a large number of nodes, such as smart meters in a smart grid, IoT (Internet of Things) network nodes, or sensor nodes in an automated factory \cite{mmtc}. An mMTC exhibits several distinct features that are uncommon in human-agent based networks. These are: uplink (UL) dominated network, sporadic activity of transmitting nodes, transmission of short sized packets,  strict latency requirements, and low data rates \cite{mmtc}. These make a conventional grant-access based legacy communication system unsuited for mMTC. Significant signaling overhead associated with grant-acknowledgement scheduling procedures results unacceptably high latency in legacy systems. It is shown in \cite{ulgf} that in 4G LTE one can have as much as 30\% grant signaling overhead to transmit a small amount of data.
Recently the UL grant-free non-orthogonal multiple access (NOMA) schemes \cite{noma,noma2,mmtc1,ulgf} have offered  promising solutions to the above mentioned problem. Grant-free NOMA allows devices/users to randomly transmit data without any complex handshaking process, while supporting massive connectivity by allocating a limited resources in a non-orthogonal manner to a massive number of machine type nodes. 

As per some empirical findings reported in  \cite{empirical}, the number of simultaneously active users at a given point of time does not exceed 10\% of the number of nodes in an mMTC network, even in its peak operation time. This inherent user sparsity in mMTC has motivated a range of Compressive Sensing (CS) based Multi User Detection (CS-MUD) techniques for active user detection (AUD), channel estimation (CE) and data detection (DD). In practice, the user activity pattern remains unchanged over several consecutive 
frames,  which is known as frame-wise sparsity in literature. 
Greedy CS-MUD algorithms like Group Orthogonal Matching Pursuit (GOMP) \cite{gomp}, weighted GOMP (WGOMP) \cite{wgomp}, and structured iterative support detection (SISD) \cite{sisd} rely on frame-wise sparsity for active user detection. For example, in \cite{ieamp,sompext} authors developed low complexity CS-MUD techniques. But these algorithms rely on assumptions which include prior knowledge about number of active users and complete knowledge of channel gain at base station respectively. However, it is not clear how these additional assumptions can be justified in practice.


Several other researchers have adopted various Bayesian methods for user detection and channel estimation \cite{sbl,map,ep} in mMTC. In \cite{ep} authors proposed expectation propagation (EP) based joint user detection and channel estimation where a computationally intractable Bernoulli-Gaussian distribution was approximated by a tractable multivariate Gaussian distribution to find an estimated \textit{posteriori} distribution of the sparse channel vector. In \cite{sbl} authors proposed Sparse Bayesian Learning (SBL) based user detection and channel estimation. Although these Bayesian approaches offer good performance, they need statistical priors like user activation probability, precise statistics of noise and channel gain.

In this paper our objective is to develop a low complexity solution. In addition, the algorithm should need no information/assumption about channel characteristics, user sparsity, activation probability, etc. Yet we want good detection estimation performance.  In this goal we attempt to find ways to engineer the spreading sequences with some special structure, and we shall exploit this structure to design fast and accurate detection-estimation algorithms.





\subsection{Contributions}
Our contributions can be summarized as follows:
\begin{itemize}
    \item We show that it is indeed possible make use of fast and accurate methods from the classical signal processing literature if we use  sinusoidal spreading sequences. The resulting signal model admits some Vandermonde structure \cite{vandermonde}. This allows us to apply fast, non-iterative  algorithms like root-MUSIC or ESPRIT \cite{rootmusic,esprit}. These subspace methods do not require prior noise/channel statistics at the BS. Furthermore, we can combine the subspace methods with some well understood information criterion, {\em i.e.} Corrected Akaike information criteria (AICc)\cite{aicc}, Bayesian Information Criteria (BIC) \cite{bic} or weighted information criteria (WIC) \cite{wic} to estimate the number of active users. Therefore, we don't need to make any assumptions on activation probability or sparsity levels. In this way, the proposed technique escapes both of the most pressing requirements of CS-MUD and Bayesian approaches. For further performance improvement, especially in scenarios with higher number of active users and high measurement noise variance, ESPRIT can be  optionally  used to initialize a Variable Projection \cite{vpn} algorithm     to solve the underlying maximum likelihood estimation problem. When initialized with ESPRIT estimates, the variable projection algorithm is known to converge in only a few iterations
    \cite{Petre97}. However, to 
    the best of our knowledge, this optional application of the variable projection algorithm is not necessary in practice.

	\item We present a new method for joint estimation of channels and data symbols. This method can exploit the additional knowledge about the signal constellations to jointly improve the channel estimation and data detection performance. Furthermore, we give conditions for reliable recovery of transmitted data symbol. In short, among estimated active users (UE) we list UEs of which data symbols cannot be recovered reliably due to poor signal to noise ratio. This may aid MAC (Media Access Control) to derive optimal power control strategies for connected devices.

    \item We carry out extensive numerical evaluations in realistic non-line of sight (NLOS) scenarios of mMTC demonstrated by 3GPP (release 9) to compare various performance metrics over several network parameters. It is demonstrated that proposed method outperforms state-of-art Bayesian algorithm \cite{ep} while maintaining less computational expenditure.
\end{itemize}

The rest of the paper is organized as follows. Section \ref{system_model} demonstrates the system model of an UL grant-free mMTC system. In Section \ref{mod_system_model}, we propose sinusoidal code as spreading sequence and transform the system model. Section \ref{esprit} delineates the methods used for model order selection and active user identification followed by Section \ref{ce_dd} where conditions of reliable data recovery is derived along with channel estimation technique. An detailed numerical investigation is carried out at Section \ref{simulations} where performance comparison and complexity analysis are discussed. We conclude and briefly discuss several future research directions in Section \ref{conclusion}. 

\section{System Model}
\label{system_model}
\begin{figure}
    \centering
    \includegraphics[width=\columnwidth]{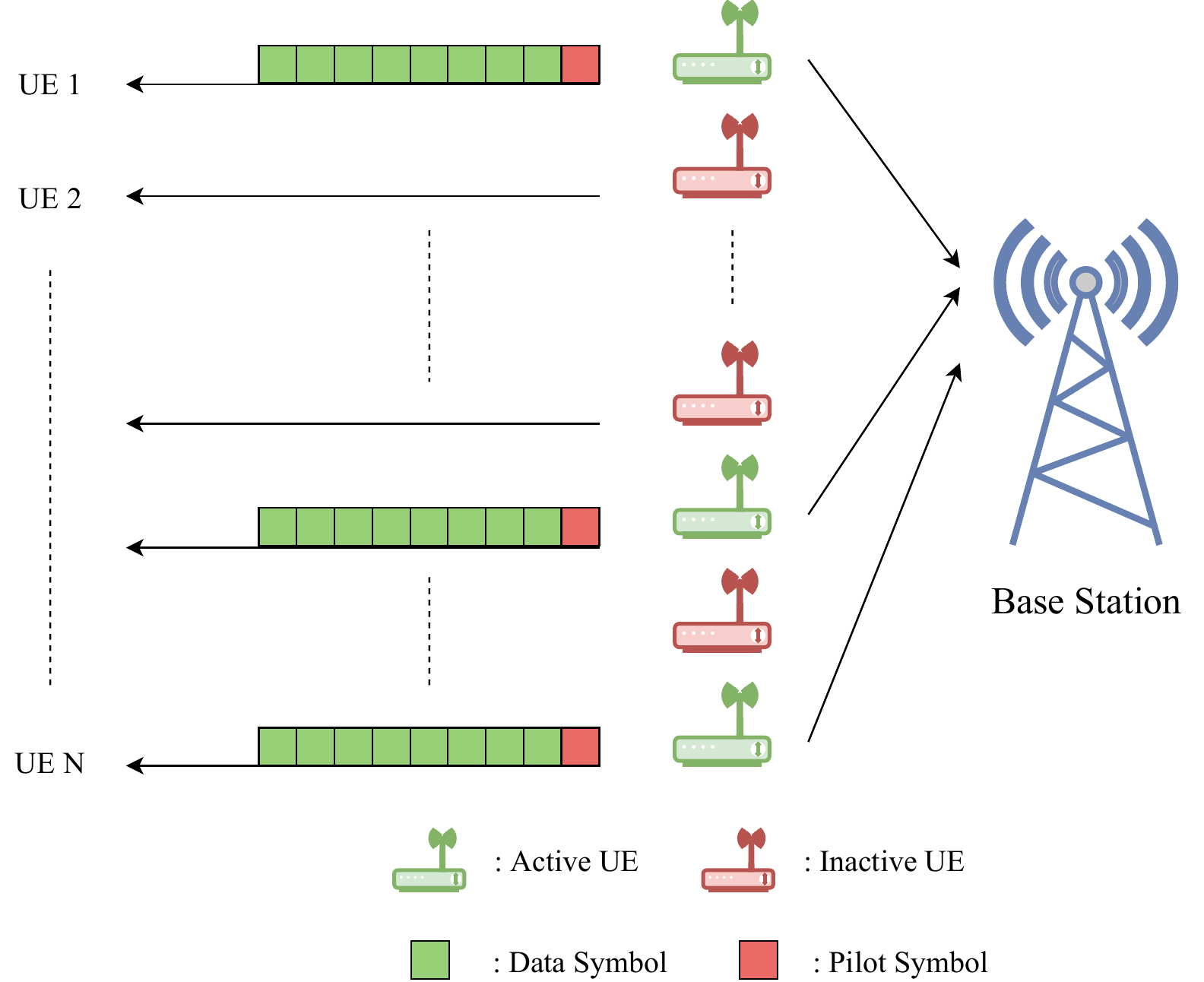}
    \caption{Uplink scenario of an mMTC cell with single BS and $N$ synchronized UEs}
    \label{mMTC}
\end{figure}

Consider a cell of an mMTC system consisting of a base station (BS) and $N$ user equipments (UE). The system allocates $M <N$ resources for creating a grant-free NOMA network consisting of the above UEs. 
The number $N$ can be rather large. Nevertheless, only a small fraction of these $N$ UEs initiate sporadic transmission at any given point of time \cite{empirical}. This fraction is commonly referred as the activation probability $p_a$ in literature. In this system each UE is allocated  its own spreading sequence. The spreading sequence allocated to user $n$ is given by an $M$ dimensional complex valued vector $\gph_n$.

During a random access opportunity, a UE transmits over $J$ time-slots. At the $j$ th time-slot, the $n$ th user spreads its complex scalar valued data symbol $\beta_{n,j}$ over $M$ orthogonal resources using its spreading sequence $\gph_n$. It forms the vector $\gph_n \beta_{n,j}$. Subsequently, it transmits the $m$ th component of $\gph_n \beta_{n,j}$ via the $m$ th orthogonal resource at $j$ th time-slot. Note that $\beta_{n,j} = 0$ for all $j$ if user $n$ is inactive.

The signal received by the BS consists of all transmissions.
Using the signal received during the random access opportunity, the BS forms the complex-valued vectors
$\{ \By_j \}_{j=1}^J$. Each $\By_j$ is an $M$ dimensional vector formed by arranging the
data received in $M$ individual resources during the $j$ th time-slot.
The vector $\By_j$ can be  written as 
\begin{align}
    \By_j &= \sum_{n \in \mathcal{N}}  \boldsymbol{\phi}_n  \beta_{n,j} h_n + \Bw_j,
    \ \ j=1,2,\ldots,J.
    \label{eq1}
\end{align}
where, $\mathcal{N}$ denotes the set of active users during the random access opportunity. In \cf{eq1} we  assume flat Rayleigh fading channels in uplink, and all $M$ orthogonal channels fall within coherent bandwidth \cite{ep}. Therefore, user $n$ experiences the same complex, scalar valued channel gain $h_n$ in all resources. 

In \cf{eq1}, $\Bw_j$ denotes the complex vector valued measurement noise. 
We assume 
that the random vectors $\Bw_1, \Bw_2, \ldots, \Bw_J$ are mutually independent, and identically distributed with a mean zero, and covariance matrix $\sigma^2 I$, where $\sigma$ is unknown.

For any active user $n$ it is assumed that $\beta_{n,1} = 1$. This is the first data symbol that acts as the pilot. Subsequently, for all $j > 1$, we assume $|\beta_{n,j}| = 1$ provided that user $n$ is active. This is true when the data symbols are drawn from some PSK constellation. This assumption on $\beta_{n,j}$ are not necessary for the code detection step of our algorithm. We use this assumption in developing a joint channel and data estimation technique. 

In the sequel we use $a_n$ to denote the index of $n$ th active user. Therefore,
\be
\mathcal{N} = 
\{ a_1, a_2 , \ldots, a_{|\mathcal{N}|} \}.
\label{setN}
\ee
Note that $| \mathcal{N}|$ denotes the cardinality of the set $\mathcal{N}$. 
Combining all received vectors over $J$ time-slots, we get the data matrix
\begin{equation}
    \bY = [\By_1 \ \By_2 \ \cdots \ \By_J].
    \nonumber
\end{equation}
Using \cf{eq1} and \cf{setN} we can write $\bY$ as 
 \begin{equation}
    \bY = \Phi\Upsilon + \bW,
    \label{father}
\end{equation}
where $\bW = [  \  \Bw_1 \ \Bw_2 \ \cdots \ \Bw_J \ ] $,
\be
\Phi = [ \ \boldsymbol{\phi}_{a_1}, \  \boldsymbol{\phi}_{a_2}, \  \ \cdots \  \ ,\boldsymbol{\phi}_{a_{\vert\mathcal{N}\vert}} \ ] ,
\label{thePhi}
\ee
and 
$\Upsilon$ is a $|\mathcal{N}| \times J$ matrix such that 
\be
\Upsilon_{n,j} = h_{a_n} \beta_{a_n,j}.
\label{upsilon}
\ee

\section{System Model with Sinusoidal Sequences}
\label{mod_system_model}
In the sequel, we work with sinusoidal spreading sequences. The $m$ th component of the spreading sequence $\gph_n$ employed by user $n$ is given by
\begin{equation}
    \boldsymbol{\phi}_n (m) = \gamma \exp( \irm 2 \pi m n/N ) \  \ 
    m \in \{0,1, \cdots M-1\}.
    \label{sin}
\end{equation}
 Since $|\beta_{n,j}| = 1$ for all $n$ and $j$, it follows that all
 UEs use the same transmit power  proportional to $\gamma^2$.

\subsection{Signal model for sinusoidal spreading sequences}
\label{newmodel}
Note that the angular frequency of the sinusoidal spreading sequence 
employed by user $n$ is given by 
\be
\omega_n = 2 \pi n/N, \ n = 0,1,\cdots,N-1.
\label{omegan}
\ee
With this, and \cf{sin}, we can write the received signal at $m$ th resource of the  $j$ th frame as 
\begin{equation}
    \By_j(m) = \sum_{n \in \mathcal{N}} \exp(\irm \omega_n m) x_{n,j} + \Bw_{j}(m),
    \label{e0}
\end{equation}
$m \in \{0,1, \cdots M-1\}$ and $j \in \{0,1, \cdots J-1\}$, where $\mathcal{N}$ denotes the set of all active user indices, and we 
write
\be
x_{n,j} =  \gamma h_n \beta_{n,j}
\label{be}
\ee
for brevity.

Let us fix an integer $l$ of our choice. Then from \cf{e0} it follows for 
any $m \in \{0,1, \cdots M-l \}$ and $j \in \{0,1, \cdots J-1\}$ that
\begin{align}
\nonumber
\Bs_{m,j}  
&:= [ \ \By_j(m) \ \By_j(m+1) \ \cdots \ \By_j(m+l-1) \ ]^{\intercal} \\
&=
\sum_{n \in \mathcal{N}} \boldsymbol{\theta}_n x_{n,j} \exp(\irm \omega_n m) + \Bw_j(m:m+l-1).
\label{e1}
\end{align}
In above $\Bw_j(m:m+l-1)$ denotes the $l$ dimensional vector made up of the $m$ th through to the  $m+l-1$ th components of $\Bw_j$,
and in addition, we define
\be
\boldsymbol{\theta}_n = [ \ 1 \ \exp(\irm\omega_n) \ \cdots \ \exp \{ \irm(l-1)\omega_n \} \ ]^{\intercal}.
\ee
From \cf{e1} it is readily verified that 
\be
\bS_j := [ \ \Bs_{0,j} \ \Bs_{1,j} \ \cdots \ \Bs_{M-l,j} \ ] = \sum_{n \in \mathcal{N}} \boldsymbol{\theta}_n \gch^{\intercal}_{n,j} + \bW_j,
\label{e3}
\ee
where 
\begin{align}
\gch_{n,j} &:= x_{n,j} [ \ 1 \ \exp(\irm\omega_n) \ \cdots \ \exp \{ \irm(M-l) \omega_n \} \ ]^{\intercal}, \\
\bW_j &= 
\left[ \ba{cccc} 
\Bw_j(0) & \Bw_j(1) & \cdots & \Bw_j(M-l) \\   
\Bw_j(1) & \Bw_j(2) & \cdots & \Bw_j(M-l+1) \\   
\Bw_j(l-1) & \Bw_j(l) & \cdots & \Bw_j(M-1) 
\ea \right].   
\end{align}

Define the $l \times l$ anti-diagonal matrix $\bK$ such that 
\[
\bK = \left [ \ba{cccc} 
0 & \cdots & 0 & 1 \\ 
\vdots &  \iddots & 1 & 0 \\ 
0 &  \iddots & \iddots & \vdots \\ 
1 & 0 & \cdots & 0 
\ea \right].
\]
For any column vector $\za$ the product $\bK \za$ is same as $\za$ flipped upside down. In particular, if $\mathrm{conj}(\za)$ denotes the complex 
conjugate of $\za$, then it is readily verified that
\[
\bK \ \mathrm{conj}(\gth_n) = \gth_n \exp \{ -\irm \omega_n (l-1) \}.
\]
Therefore, \cf{e3} yields that
\[
\bK \ \mathrm{conj}(\bS_j) = 
\sum_{n \in \mathcal{N}} \boldsymbol{\theta}_n \gch^{*}_{n,j} \exp \{ -\irm \omega_n (l-1) \} + \bK  \ \mathrm{conj}(\bW_j),
\]
where $\gch^{*}_{n,j}$ denotes the conjugate transpose of $\gch_{n,j}$. By combining the last equality with \cf{e1} it follows that
\be
\bar{\bS}_j := 
[ \ \bS_j \ \ \bK \ \mathrm{conj}(\bS_j) \ ] 
=
\sum_{n \in \mathcal{N}} \boldsymbol{\theta}_n \bar{\gch}^{*}_{n,j} +   \bar{\bW}_j,
\label{e2}
\ee
where
\begin{align}
\begin{split}
\bar{\gch}_{n,j} &= [ \ \gch_{n,j}^{\intercal} \ \ \gch^{*}_{n,j} \erm^{ -\irm \omega_n (l-1)} \ ]^*,
\\ 
\bar{\bW}_j &= [ \ \bW_j \ \ \ \bK  \ \mathrm{conj}(\bW_j) \ ].
\end{split}
\end{align}
Now considering \cf{e2} for $j=0,1,\ldots,J-1$, we get 
\be
\overline{\bS} = [ \ \bar{\bS}_0 \ \bar{\bS}_1 \ \cdots \ \bar{\bS}_{J-1} \ ] =
\sum_{n \in \mathcal{N}} \boldsymbol{\theta}_n \gps^{*}_{n} +   \overline{\bW},
\label{son}
\ee
where
\begin{align}
\begin{split}
\gps_{n} &= [ \ \bar{\gch}_{n,0}^* \ \ \bar{\gch}^{*}_{n,1} \ \ \cdots \ \ \bar{\gch}_{n,J-1}^* \ ]^*,
\\ 
\overline{\bW} &= [ \ \bar{\bW}_0 \ \ \bar{\bW}_1  \ \ \cdots \ \ \bar{\bW}_{J-1} \ ].
\end{split}
\end{align}
Given the signal received at the BS, and a suitable integer $l$ of our choice, we can readily form the data matrix $\overline{\bS}$ 
via the first equalities in the equations 
\cf{e1}, \cf{e3}, \cf{e2} and \cf{son}. 
The matrix $\overline{\bS}$ has $l$ rows and $2J(M-l+1)$ columns. 
The rank of it's noise-free part, which is given as
\[
\overline{\bS} - \overline{\bW} =  \sum_{n \in \mathcal{N}} \boldsymbol{\theta}_n \gps^{*}_{n},
\]
can not be more than   $|\mathcal{N}|$.   
Therefore, if we take a large enough $l$ so that $l$ is always larger than $|\mathcal{N}|$, then $\overline{\bS} - \overline{\bW}$ will be of
 rank $|\mathcal{N}|$.  This property can be exploited by subspace algorithms \cite{stoica} . These algorithms work
by obtaining a low rank approximation of $\overline{\bS}$ via its singular value decomposition. The details are discussed below.


\section{Fast Subspace Based AUD}
\label{esprit}
\subsection{Estimation of Number of Active Users}
Note that the column-space of  of the noise-free data $\overline{\bS} - \overline{\bW}$ is spanned by the vectors 
$\gth_{n},  n \in \mathcal{N}$. Subspace algorithm estimates this subspace via the eigen decomposition of $\overline{\bS}$   \cite{stoica}. Subspace algorithms like MUSIC \cite{rootmusic} or ESPRIT \cite{esprit} require the knowledge of number of signals to be specified as a \textit{priori}. This is also the case for most of the CS-MUD algorithms where this information is decoded as prior sparsity level \cite{ieamp}. However, there exist sparsity-blind greedy algorithms which assume the availability of perfect channel knowledge or noise statistics at BS \cite{sompext,ulgf_2}. In Bayesian approaches, this information is supplied to the algorithms as user activation probability $p_a$ which is derived from empirical studies \cite{ep}. In this work, however, we assume none of these information are available to BS. Specifically, we employ an information-criteria aided model order selection technique to find the number of active users in a random access opportunity.

In the sequel  $\hat{\mathcal{N}}$ denotes the set of estimated active user indices. 
 Hence, $\vert\hat{\mathcal{N}}\vert$ denotes the number of estimated active users.  
Given the data matrix $\overline{\bS}$ an information-criterion aided technique estimates 
$\vert\hat{\mathcal{N}}\vert$ as
\begin{equation}
\vert\hat{\mathcal{N}}\vert   =  \arg \min_k \ -\log f(\overline{\bS},k) + \mathcal{W}_k
 \label{ic}
\end{equation}
where $\mathcal{W}_k$ is a bias correction term that depends on the information criterion being used, 
and   as shown in  \cite{ic}, the log-likelihood function in \eqref{ic} is given by
\begin{equation}
    \log f(\overline{\bS},k) = (l-k)\mathcal{\mathcal{P}} \ \ln 
    \left \{ \frac {\displaystyle \prod_{i=k+1}^l (\varsigma_i^2/\mathcal{\mathcal{P}})^{\frac{1}{l-k}}}{\frac{1}{l-k}\displaystyle \sum_{i=k+1}^{l} \varsigma_i^2/\mathcal{\mathcal{P}}} \right \},
    \label{loglikelihood}
\end{equation}
where $\mathcal{P}$ is the number of columns in $\overline{\bS}$ \textit{i.e:} 
\[
\mathcal{P} = 2J(M-l+1),
\]
and $\{\varsigma\}_{i=1}^{l}$ denote the $l$ non-zero singular values of the $l \times \mathcal{P}$ matrix $\overline{\bS}$.
In particular, we assume without any loss of generality that
\[
\varsigma_1 > \varsigma_2  > \cdots > \varsigma_l.
\]

Several different information criteria can be used to determine how $\mathcal{W}_k$ depends on $k$ \cite{ic,wic,icstudy,iccomp}. 
For example one can use the classical information criterion by Alaike \cite{Akaike1998}, for which  $\mathcal{W}_k = k(2l-k)$ 
\cite{ic}. However, it is wellknown that AIC leads to overestimation of $|\hat{\mathcal{N}} |$. From that perspective it is better 
to use the Bayesian Information Criteria (BIC) proposed in \cite{bic} for which
\begin{equation}
    \mathcal{W}_k = \frac{1}{2}k(2l-k)\log \mathcal{P}.
    \label{bias_correction}
\end{equation}

Using the singular value decomposition of $\overline{\bS}$ one can readily calculate the cost in \cf{ic} via 
\cf{loglikelihood} and \cf{bias_correction} for different values of $k$. Therefore it is rather straightforward to estimate
$| \hat{\mathcal{N}}|$.

\subsection{Finding the Active User Indices}
\label{esprit_desc}
Among many suitable candidates of subspace algorithms such as MUSIC, root-MUSIC, in this work, we employ ESPRIT \cite{esprit}. ESPRIT has been recommended as a first choice in frequency estimation problem in \cite{stoica}. In particular, the 
data model \cf{son} leads to ESPRIT with forward-backward averaging which has been shown to have substantially 
improved statistical performance figures \cite{feb_esprit}. This process yields the set of angular frequencies of sinusoidal
 spreading sequences $\{\hat{\omega}_n\}_{i=1}^{\vert\hat{\mathcal{N}}\vert}$ used by the active users. 
The main steps of the  ESPRIT algorithm is as follows:
\begin{enumerate}
\item Construct $\Theta = [\Bu_1 \ \Bu_2 \ \dots \ \Bu_{\vert\hat{\mathcal{N}}\vert}]$,  where $\Bu_k$ denotes the
unit norm left singular vector of $\overline{\bS}$ associated with its $k$ th largest singular value $\varsigma_k$.

\item  Solve the equation 
\[
\Theta(1:l-1,:)  \ \bQ = \Theta(2:l,:) 
\]
for $\bQ$ in least squares sense. 
\item Compute eigenvalues $\{\nu_n\}_{i=1}^{\vert\hat{\mathcal{N}}\vert}$ of $\bQ$.

\item Estimate the frequencies $\{\hat{\omega}_n\}_{i=1}^{\vert\hat{\mathcal{N}}\vert}$ as 
\[
\hat{\omega}_n = \arg( \nu_n), \psp n = 1,2,\ldots, | \hat{\mathcal{N}} |.
\] 
\end{enumerate}

We use $\hat{a}_n$ to denote the $n$ th estimated active user index. In other words,
\[
\hat{\mathcal{N}}
=
\{ 
\hat{a}_1, 
\hat{a}_2, 
\ldots,
\hat{a}_{ | \hat{ \mathcal{N} } | }
\}.
\]
To find $\hat{a}_n$, we use the definition of $\omega_n$ given in 
\cf{omegan},  to get
\[
\hat{a}_n = \mathrm{round} \left(  \frac{N \hat{\omega}_n }{2 \pi} \right).
\]

\subsection{Maximum Likelihood Estimate of Active User Indices}

If the noise variance in \cf{son} is large then it one may optionally improve the accuracy of the ESPRIT estimates by applying
the Gaussian maximum-likelihood method \cite{Petre97}, 
which requires us to solve, 
\begin{equation}
\underset{ \Upsilon,  \  \{ \hat{\omega}_n \}_{n=1}^{ | \hat{\mathcal{N}} |  }}{ \mathrm{minimize}} \
 \Vert  \bY - \Phi\Upsilon \Vert_F^2,
    \label{mother}
\end{equation}
see \cf{father} and \cf{sin}  to recall how $\Phi$ depends on
$ \{  \hat{\omega}_n \}_{n=1}^{ | \hat{\mathcal{N}} |  }$. 
Noting that, \eqref{mother} is quadratic in $\Upsilon \in \mathbb{C}^{\vert\mathcal{N}\vert \times J}$, we can use linear least square to minimize \eqref{mother} with respect to $\Upsilon$ which yields
\begin{equation}
\arg \min_{\Upsilon}    \Vert  \bY - \Phi\Upsilon \Vert_F^2
= (\Phi^*\Phi)^{-1}\Phi^*\bY,
 \label{lls}
\end{equation}
and consequently it follows that
\begin{equation}
\min_{\Upsilon}    \Vert  \bY - \Phi\Upsilon \Vert_F^2
= 
|| \bY^* \{   \bI - \Phi (\Phi^*\Phi)^{-1}\Phi^* \} \bY ||_F^2.
 \label{lls1}
\end{equation}
From \cf{thePhi} note that $\Phi$ depends on the spreading sequences, which by \cf{sin} depends on the frequencies.  Hence
the reduced cost function in the right hand side of \cf{lls} depends on the frequencies.  Hence the maximum likelihood
estimate of the frequencies can be obtained by solving 
\begin{equation}
\underset{  \{ \hat{\omega}_n \}_{n=1}^{ | \hat{\mathcal{N}} | }}{ \mathrm{minimize}} \
|| \bY^* \{   \bI - \Phi (\Phi^*\Phi)^{-1}\Phi^* \} \bY ||_F^2.
 \label{mother1}
\end{equation}
The cost function in \cf{mother1}  is  non-convex in $\hat{\omega}_n, \ n \in \hat{\mathcal{N}}$.  Nevertheless, \cf{mother1} turns out
to be a variable projection problem, \cite{vpn} for which well known algorithms exist \cite{vpn}. 
These algorithms are wellknown to converge very quickly when initialized at a point near the global  optimum. 
For our problem we can use the  ESPRIT estimates  $\{\hat{\omega}_n\}_{n \in \hat{\mathcal{N}} }$ to initialize  
the variable projection algorithm.  
As will be reiterated in later section, the algorithm converges within $2\sim 3$ iterations when $p_a<0.25$. Upon convergence, we update the frequency estimates $\{\hat{\omega}_n\}_{n \in \hat{\mathcal{N}} }$  by the corresponding 
ML estimates and find the set of active user indices $\hat{\mathcal{N}}$ as mentioned in previous section.


\section{Channel Estimation and Data Detection}
\label{ce_dd}
In this section, we delineate the methods of channel estimation and data detection. In contrast with existing works where only signal received in the pilot frame is used, we use the entire received signal matrix $\bY \in \mathbb{C}^{M \times J}$ for channel estimation. The role of pilot symbol is to estimate the phase angle of estimated complex channel gains $\{h_i\}_{i=1}^{\vert\hat{\mathcal{N}}\vert}$ accurately as will be shown in following section.
\subsection{Channel Estimation}
\label{ce}
Using the detected active user indices  we construct the 
$M \times | \hat{ \mathcal{N} } |$  matrix $\hat{\Phi}$ 
such that
\begin{equation}
[\hat{\Phi}]_{m,n} = 
\gamma \exp( \irm2\pi m \hat{a}_n / N ).
\label{scn}
\end{equation}
Using $\hat{\Phi}$  we can find an estimate of $\Upsilon$,  see \cf{lls}, 
\begin{equation}
    \hat{\Upsilon} = (\hat{\Phi}^*\hat{\Phi})^{-1}\hat{\Phi}^*\bY.
    \label{lls1}
\end{equation}
Note that $\Tilde{\Upsilon}\in\mathbb{C}^{\hat{\vert\mathcal{N}\vert}\times J}$. 

In the system considered herein  uses  a
 a $L$-PSK constellation. Define the set
\be
\mathcal{A} = 
\{0,1,2,\cdots \ L-1\}.
\ee
Hence  $\beta_{a_n,j}$ must be of the form 
\begin{equation}
    \beta_{a_n,j} = \exp(  \irm 2\pi q_{a_n,j} / L)
     \label{qpsk}
\end{equation}
where each  $q_{a_n,j} \in  \mathcal{A} $. The data detection problem requires us to estimate $q_{\hat{a}_n,j}$ for all
$\hat{a}_n \in \hat{\mathcal{N}}$ and $j = 2, 3, \ldots, J$. We remind the readers that the pilot symbols equal to unity, 
{\em i.e.}
\be
\beta_{\hat{a}_n,1} = 1, \psp  \Leftrightarrow \psp q_{\hat{a}_n,1} = 0
\label{jis1}
\ee
for all $\hat{a}_n \in \hat{\mathcal{N}}$.

Now recalling \cf{upsilon} and using \cf{qpsk} we can write 
\begin{align}
\Upsilon_{n,j} 
&= 
h_{a_n} 
\beta_{ a_n,j} 
= 
|h_{a_n}|  
\exp \left (
\irm 
\left \{
\zeta_{ a_n }
+
\frac{2\pi}{L}q_{a_n,j}  
\right \}
\right )
\label{neely}
\end{align}
where $\zeta_{a_n}$ is the phase angle of channel gain \textit{i.e:}
$h_{a_n} =\vert h_{a_n} \vert \exp(\irm\zeta_{a_n})$. 
Taking natural logarithm at both sides of \eqref{neely} we get
\begin{equation}
    \ln{( \Upsilon_{n,j})} = \ln(\vert h_{a_n} \vert) + \irm\left(\zeta_{a_n} + \frac{2\pi}{L}q_{a_n,j}\right),
    \label{ln}
\end{equation}
so that 
\be
\mathrm{Re} \{   \ln{( \Upsilon_{n,j})}  \}
= \ln(\vert h_{a_n} \vert) ,
\label{real_stuff}
\ee
\be
L
\mathrm{Im} \{   \ln{( \Upsilon_{n,j})}  \}
=
L  \zeta_{a_n} +  2\pi q_{a_n,j}.
\ee
Since each $q_{a_n,j} \in \{ 0,1,2,\ldots,L-1\}$, it follows that
\be
\mathrm{mod} (
L \mathrm{Im} \{   \ln{( \Upsilon_{n,j})}  \}  \ , \  2 \pi
)
=
\mathrm{mod} (
L  \zeta_{a_n}  \ , \ 
 2\pi 
 ).
\label{imag_stuff}
\ee
Next we replace $\Upsilon$ by its estimate $\hat{\Upsilon}$ in \cf{lls1}. 
The estimate $\hat{\Upsilon}$ will not be error-free. We propose to 
reduce the effect of the estimation error by averaging. 
Using \cf{real_stuff} we estimate $|h_{\hat{a}_n}|$ 
for each $\hat{a}_n \in \hat{ \mathcal{N}}$  as
\be
\vert \hat{h}_{\hat{a}_n} \vert
=
\frac{1}{J} \sum_{j=1}^J
\mathrm{Re} \{   \ln{(  \hat{\Upsilon}_{n,j})}  \}.
\label{est_real_stuff}
\ee
Similarly,  using \cf{imag_stuff} we estimate  $\mathrm{mod} (L  \zeta_{ \hat{a}_n}  \ , \ 2\pi )$ by 
\be
\bar{\zeta}_{ \hat{a}_n} = 
\frac{1}{J} \sum_{j=1}^J
\mathrm{mod} (
L \mathrm{Im} \{   \ln{( \Upsilon_{n,j})}  \}  ,  2 \pi).
\label{est_imag_stuff}
\ee
Since $\bar{\zeta}_{ \hat{a}_n}$ is an estimate of $\mathrm{mod} (L  \zeta_{ \hat{a}_n}   ,  2\pi )$ ,
we have $L$ possible candidates which can be an estimate of  $\zeta_{ \hat{a}_n}$. These are the elements of the set
\[
\mathcal{J} = 
\left \{ \bar{\zeta}_{ \hat{a}_n} + \frac{2 \pi k}{L} \ : \ k \in \mathcal{A} \right \}.
\]
To identify the correct candidate from the above set we make use of \cf{jis1} and \cf{neely}, which yield
\begin{align}
\Upsilon_{n,1} 
&= h_{a_n} =
|h_{a_n}|  
\exp(  \irm  \zeta_{ a_n }).
\label{neely_j1}
\end{align}
Equation \cf{neely_j1} reveals that  $\hat{\Upsilon}_{n,1}$ is, in fact, an estimate of  $h_{\hat{a}_n}$.
But this estimate being based on only one element of $\hat{\Upsilon}$, is more noisy than our proposed 
estimate based on the averaged statistics in \cf{est_real_stuff} in \cf{est_imag_stuff}.
Nevertheless, we can use  $\hat{\Upsilon}_{n,1}$  to identify the correct 
candidate from $\mathcal{J}$. In particular we estimate $\zeta_{\hat{a}_n}$ by
\be
\hat{\zeta}_{\hat{a}_n} = \arg \min_{\zeta \in \mathcal{J}}  
\
\left |
\erm^{\irm \zeta} -
\frac{\hat{\Upsilon}_{n,1}   }{ |\hat{\Upsilon}_{n,1}   | }
\right |.
\label{zeta_estimate}
\ee

\subsection{Data Detection}
Our data detection method is based on \cf{neely}, which yields
\begin{align}
\frac{\Upsilon_{n,j} }{|\Upsilon_{n,j}|} \erm^{-\irm  \zeta_{ a_n } }
= 
\exp( \irm 2\pi q_{a_n,j} /L).
\label{neely_2}
\end{align}
Motivated by \cf{neely_2}, and using the estimates $\hat{\Upsilon}_{n,j}$ and
$\hat{\zeta}_{\hat{a}_n}$  (obtained in \cf{zeta_estimate}) we propose to
estimate $q_{\hat{a}_n,j}$ as 
\be
\hat{q}_{\hat{a}_n,j}
=
\arg \min_{q \in \mathcal{A}}
\
\left |
\frac{ \hat{\Upsilon}_{n,j} }{| \hat{\Upsilon}_{n,j}|} 
\exp(-\irm  \hat{\zeta}_{ \hat{a}_n } )
-
\exp(  \irm  2\pi q  / L )
\right |.
\label{data_detection}
\ee



\subsection{Condition for Reliable Data Detection}
\label{condition}

An UE located near the cell boundary  experiences high path-loss, and have low signal to noise ratio (SIR).  It
is wellknown that the detection-estimation algorithms have difficulties in 
detecting such users experiencing deep fading. Nevertheless, in our simulation
study we have observed that the proposed user detection method is often capable of
detecting such low-SNR UEs, but the data detection method outlined in the 
previous sub-section may not produce accurate results.  It is of significant 
interest for the system designer to be able to detect such users. Because this 
allows the BS to send an ARQ like signal to this users, and those users
can use a higher transmit power in their next random access attempt. Such ARQ
like mechanisms can greatly improve the overall network performance. In this 
section we briefly discuss how such low SNR users could be identified within the
framework proposed herein.





\begin{figure}[t]
    \centering
    \subfloat[ \label{Ra}]{\includegraphics[width=\columnwidth]{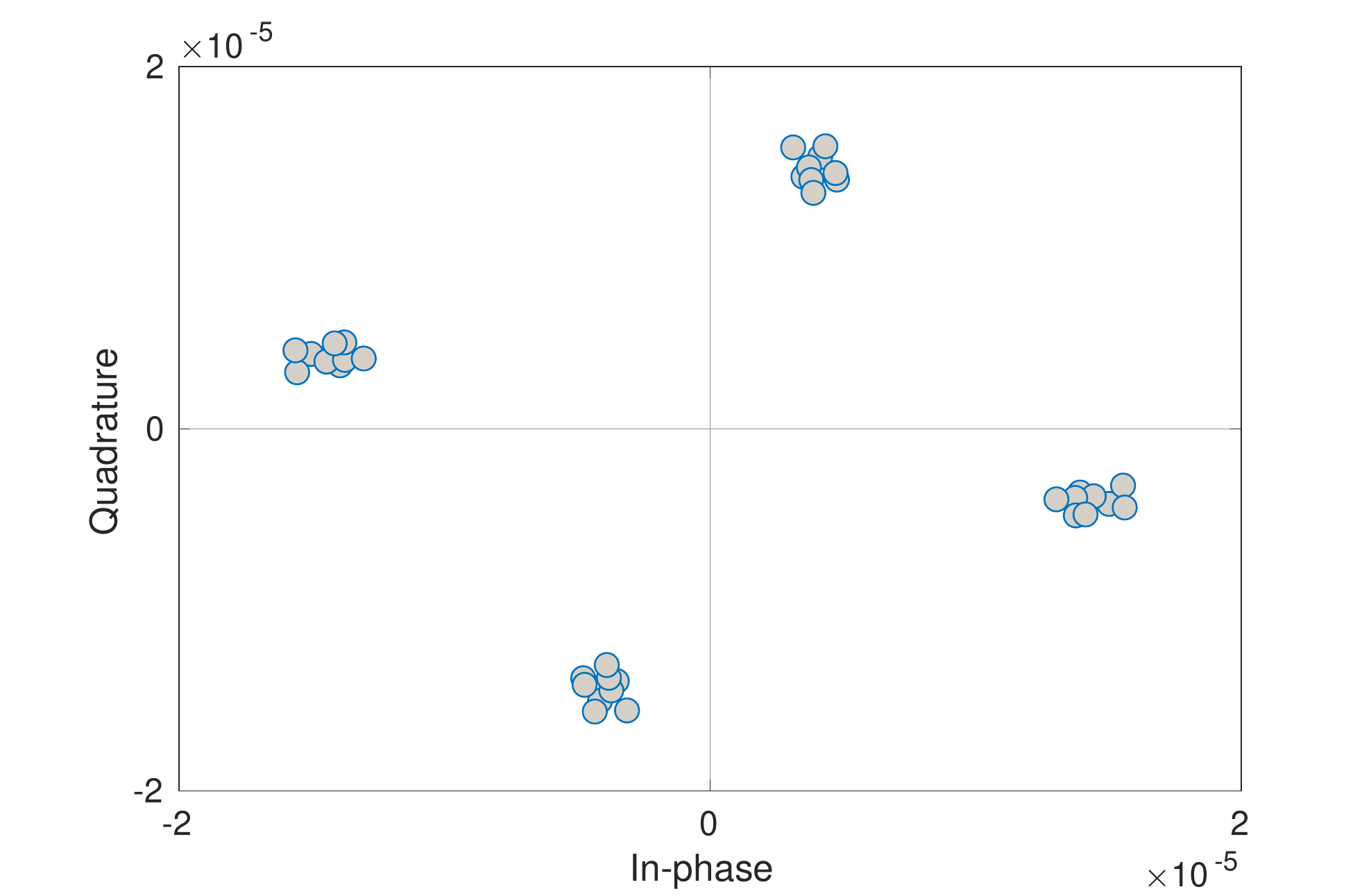}}
    
    \subfloat[ \label{Rb} ]{\includegraphics[width=\columnwidth]{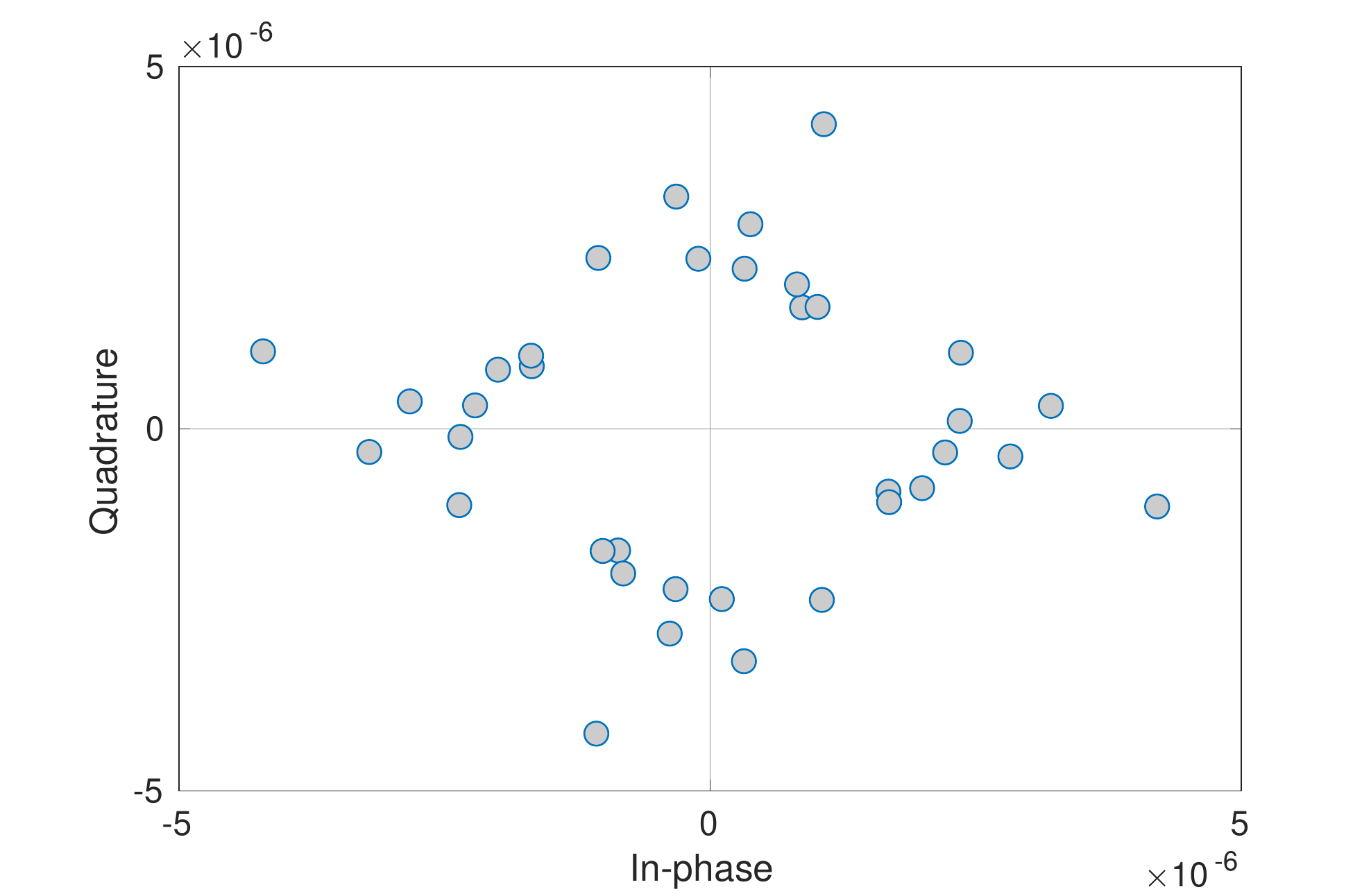}}
    \caption{(a) Received symbols of reliable UE (b) Received symbols of a non-reliable UE}
    \label{reliability}
\end{figure}

We use Figure \ref{reliability} to examine why data detection becomes difficult
for a low SNR user. In Figure~\ref{Ra} we show the typical scatter plot
of  $\hat{\Upsilon}_{n,j}$ for a fixed $n$ as $j$ varies from $1$ to $J$. 
In this example $L = 4$.
Recall
from \cf{neely} that $|\Upsilon_{n,j}| = |h_{a_n}|$ for all $j$, while the phase of
$\Upsilon_{n,j}$ can have only $4$ different values. In particular, a
scatter plot of $\Upsilon_{n,j}$ should look like the $4$-PSK 
constellation rotated by $\zeta_{a_n}$.  The plot in  Figure~\ref{Ra} is consistent
with this expectation with exception that we see the  points 4 distinct clusters 
instead of exactly 4 points. This is not a surprise as the deviation from the cluster
centers are caused by the estimation errors in $\hat{\Upsilon}_{n,j}$ contributed
by the measurement noise in the observed data. In this case the spread of the 
points caused by the measurement noise is small compared to the average value of
$|\hat{\Upsilon}_{n,j}|$ which determine the radius of the circle on which the 
cluster center lie. 

The situation becomes worse for another user with a scatter plot shown
Figure~\ref{Rb}. This user is experiencing much higher path-loss, and thus has a
significantly smaller averaged $|\hat{\Upsilon}_{n,j}|$. 
This significantly shrinks the circle
on which the cluster centers lie. Consequently, the inter-cluster separation is of
the same order as the perturbation of the constellation points due to noise. At this
stage the data detector may not be able to provide perfect data detection results.

The above example reveals that we can have perfect data detection only if the 
averaged $|\hat{\Upsilon}_{n,j}| $ is large compared to the perturbation due to noise.  
From \cf{real_stuff} we note that  $|\hat{h}_{\hat{a}_n}|$ is the 
 average value of $|\hat{\Upsilon}_{n,j}| $.  One simple way to estimate the spread
 caused by the noise would be calculate the corresponding standard deviation
 \be
 \eta_{\hat{a}_n}
 :=
 \sqrt{
 \frac{1}{J} \sum_{j=1}^J  
 \left (
 \mathrm{Re} \{  \mathrm{ln} ( \hat{\Upsilon}_{n,j} ) \}  
 -
 |\hat{h}_{\hat{a}_n}|
 \right)^2.
 }
 \label{the_eta}
 \ee
 Now the distance between the neighboring cluster centers for the estimated
 active user index is  $2 |\hat{h}_{\hat{a}_n}| \sin(\pi/L)$. This distance must 
 be more than twice the estimation error spread for reliable data detection. 
 Assuming normal distributed estimation error, we  take the spread to be 
 $5 \eta_{\hat{a}_n}$ with $99.9\%$ confidence. Hence to ensure $99.9\%$ 
 confidence in data detection we need
 \begin{align}
 \nonumber
 &
 2 |\hat{h}_{\hat{a}_n}| \sin(\pi/L) > 2 \times 5  \eta_{\hat{a}_n}
 \psp
 \Leftrightarrow
 \\
 &
|\hat{h}_{\hat{a}_n}| / \eta_{\hat{a}_n}/ > \lambda = 
5/\sin(\pi/L).
 \label{rell}
 \end{align}
 To increase the confidence level we need to increase $\lambda$.  Note that
 $\lambda$ increases if either of $L$ or the noise power increases. 
 In practice, one can implement a protocol where the BS proceeds with data 
 detection if \cf{rell} holds. Otherwise, BS can notify the user to transmit the data 
 once more with higher transmit power.



\subsection{Summary of fast subspace based AUD, CE and DD}
In this section, we summarize the steps of the proposed  algorithms.
The input of the algorithm is the received signal matrix $\bY\in\mathbb{C}^{M\times J}$ as in \eqref{father}. With these, we form the data matrix $\overline{\bS}$ as in \eqref{son}. Then the proposed method carries out following steps in sequel:
\begin{enumerate}
    
    \item Computes singular value decomposition (SVD) of $\overline{\bS}$
    
    \item Estimates $\vert\hat{\mathcal{N}}\vert$  via \eqref{ic}. For that it
    must calculate the log-likelihood  as in \eqref{loglikelihood} and  the 
    bias correction term in \eqref{bias_correction} for each competing model order $k$.

    \item Estimates angular frequencies $\{\hat{\omega}_n\}_{n=1}^{\vert\hat{\mathcal{N}}\vert}$ using a subspace algorithm.  In the sequel we have used forward-backward  ESPRIT outlined in  Section \ref{esprit_desc}.  One may optionally use the ESPRIT estimates to initialize a variable projection algorithm 
    \eqref{mother1}, and find Gaussian ML estimates, which are more accurate.
    
    \item Using the frequency estimates obtained in previous stage, it determines the set of estimated active user indices $\hat{\mathcal{N}}$, and find $\hat{\Upsilon}$, see \cf{lls1}.

    \item For each $n=1,2,\ldots,\hat{\mathcal{N}}$ 
    \bit
    \item
    Finds $|\hat{h}_{\hat{a}_n}|$
    in \cf{est_real_stuff}, $\hat{\zeta}_{\hat{a}_n}$ in \cf{zeta_estimate}, 
    and $\eta_{\hat{a}_n}$ in \cf{the_eta}. 
    
    \item 
    If $\eta_{\hat{a}_n}/|\hat{h}_{\hat{a}_n}| > \lambda$ then carries out data detection for $j=2,3,\ldots,J$ as per \cf{data_detection}. Otherwise  requests the user to retransmit with  higher transmit power.
   
   \eit
   \end{enumerate}

\section{Numerical simulation studies}
\label{simulations}
\subsection{Simulation Setup}
We consider an mMTC scenario  with $N = 128$ UEs. These are randomly deployed within the cell of radius 200 m. 
An UE is at least 1m away from the BS. The variance $\tau_n$ (in dB) of $h_n$ is modeled using NLOS model
in 3GPP (release 9):
\begin{equation}
    \tau_n = -128.1 - 36.7\log_{10}(d_n),
\end{equation}
where $d_n$ is the distance between $n$ th UE and BS in km. For receiver noise, the power spectral density is set at $-170$ dBm/Hz, and the transmission bandwidth is set at 1 MHz. A random access opportunity consists of $J=9$ frames, the first ($j=1$) of which is used to transmit pilot symbol. The pilot symbol  $\beta_{n,1}=1$ for all active UE. The PSK constellations use $L =4$. This setup is identical to that used in \cite{ep}.  

The expectation-propagation (EP) approach proposed in \cite{ep} is used as the benchmark for performance comparison. 
This choice is well justified.  In  \cite{ep} EP has been shown outperform all major existing methods by a 
comprehensive margin. 

As performance metrics, we consider Missed Detection Rate (MDR), Net Symbol Error Rate (NSER) and Root Mean Squared Error (RMSE) of CE. MDR refers to the rate of UEs failing to get recognized by candidate algorithms during AUD. On the other hand, NSER is the symbol error rate experienced by active UEs. These performance metrics are evaluated as a function of sequence length $M$, user activation probability $p_a$ and transmit power. Each point in the plots shown in the sequel are based on 10000 independent Monte-Carlo simulations. Performance of EP algorithm is evaluated using both random and sinusoidal spreading codes which are denoted as EP with RC and EP with SC, respectively. The performance of ESPRIT algorithm alone is denoted as ESPRIT whereas the performance of ESPRIT initialized variable projection based estimation is denoted as ESP-VPN. Unless specified otherwise,  $M = 64$, $p_a = 0.1$, and all UEs transmit with $20$ dBm. For ESPRIT we need to choose the integer $l$
in \cf{e1}. In our simulations we vary $l$ depending on $M$ as follows
\bc
\begin{tabular}{|c||c|c|c|c|c|c|}
\hline
$M$ & 32 & 48 & 64 & 80 & 96 & 112 \\
\hline
$l$ & 20 & 40 & 50 & 60 & 78 & 90 \\
\hline
\end{tabular}
\ec
These values are chosen to optimize the estimation performance.



\subsection{Simulation Results}

\begin{figure}[t]
    \centering
    \subfloat[\label{Ma}]{\includegraphics[width=\columnwidth]{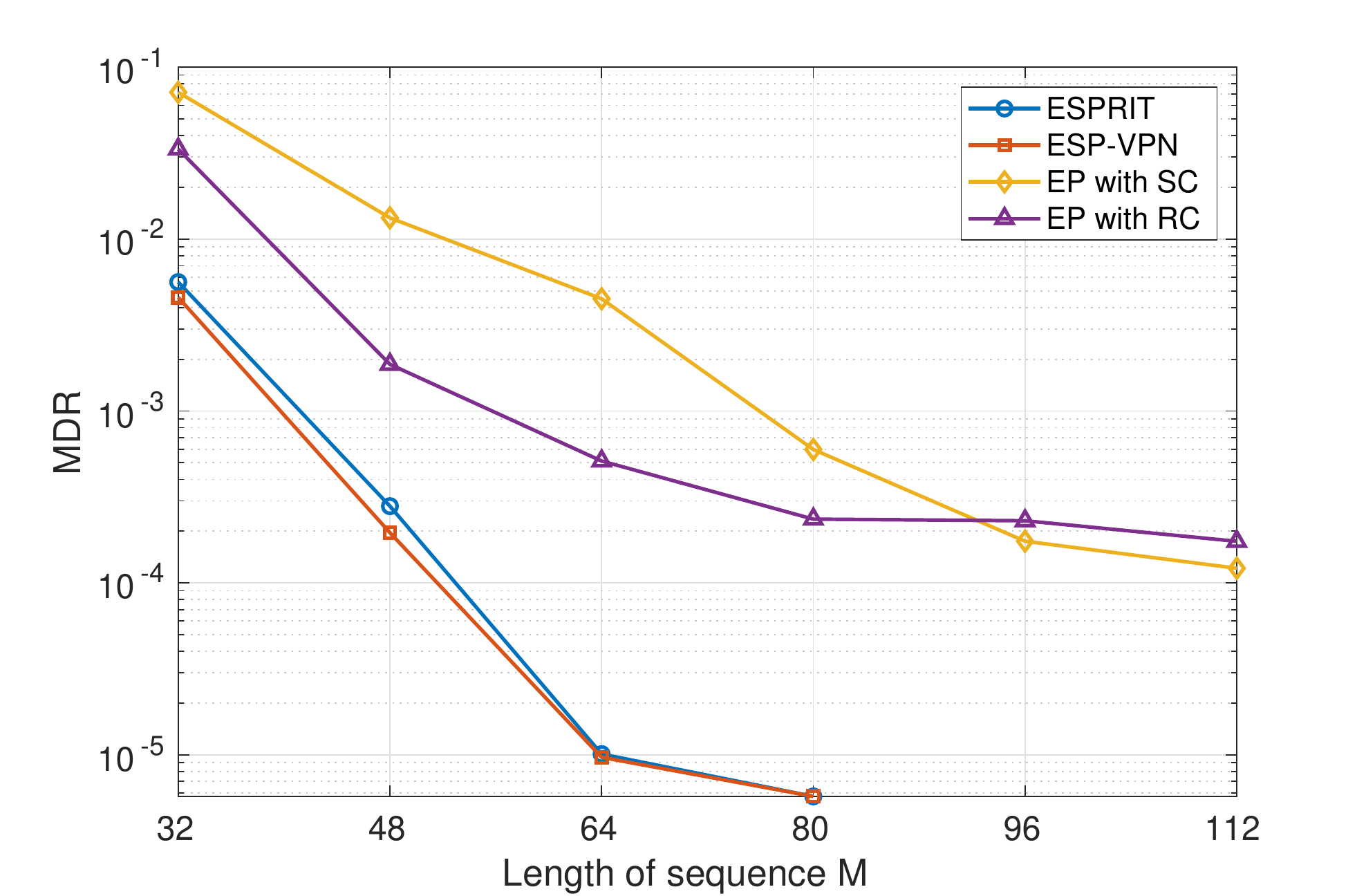}}
    
    \subfloat[\label{Mb}]{\includegraphics[width=\columnwidth]{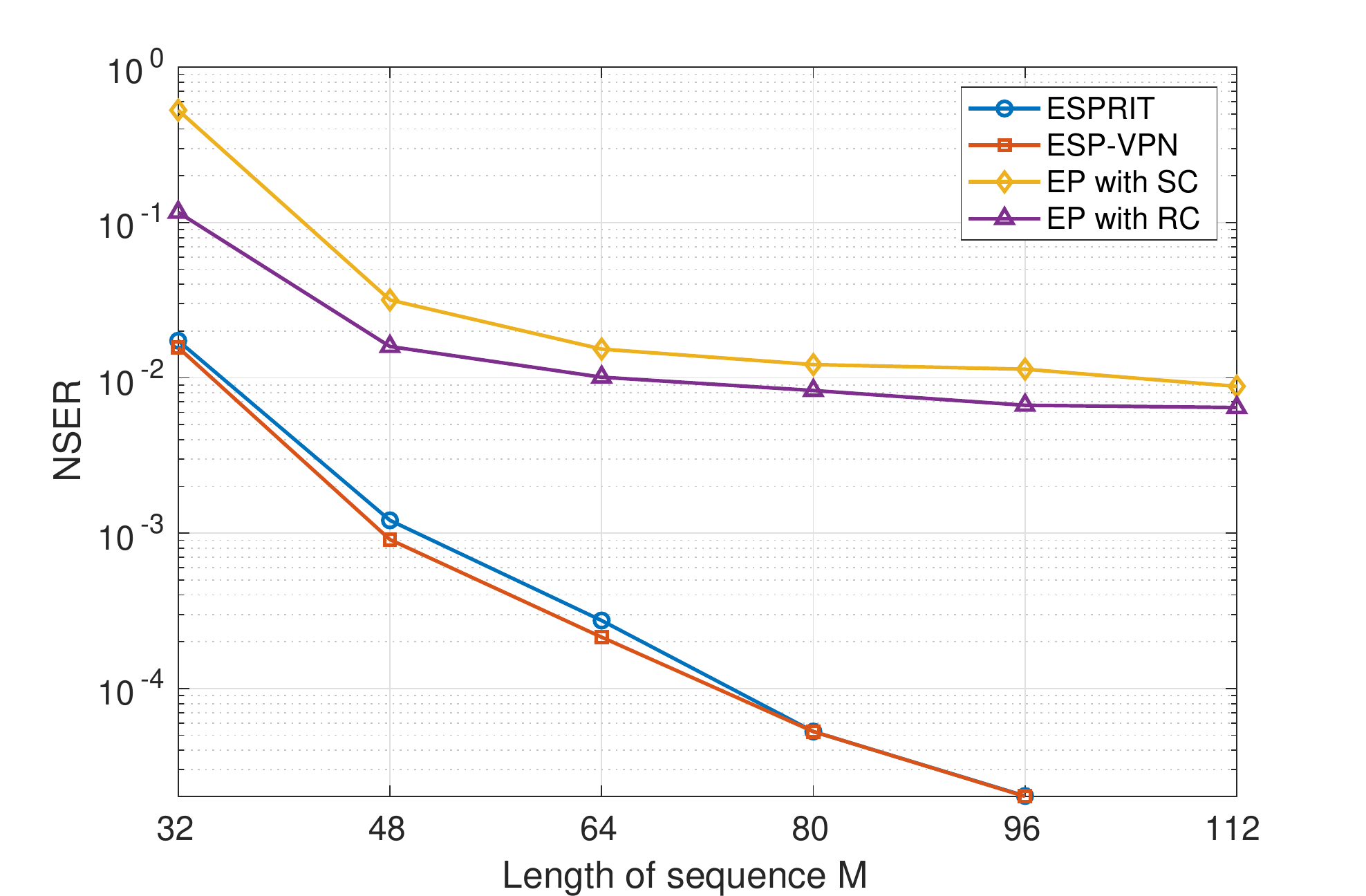}}
    
    \subfloat[\label{Mc}]{\includegraphics[width=\columnwidth]{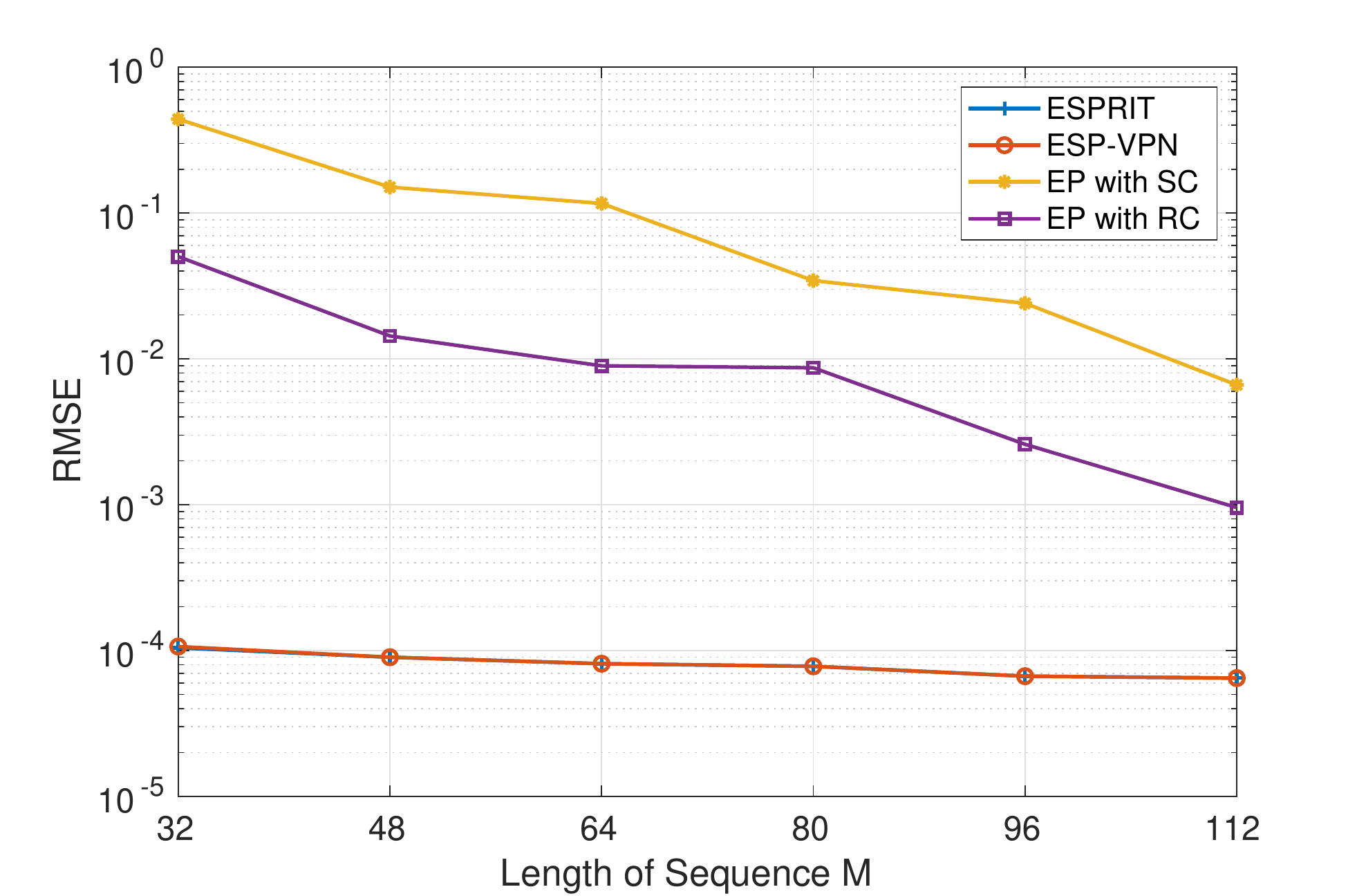}}
    \caption{(a) MDR (b) NSER (c) RMSE  of different methods as functions of $M$.}
    \label{M}
\end{figure}




Figure~\ref{M} plots the MDR, NSER and RMSE performance of different
algorithms where $M$ is varied from $32$ to $112$. In this study we fix transmit power
to  $20$ dBm and $p_a = 0.1$. As can be seen in Figure~\ref{M}, the proposed algorithm achieves significant performance improvement in all performance metrics over the EP algorithm. ESP-VPN offers slight performance improvement for $M<64$. Unlike EP, channel estimation accuracy of the proposed method  does not vary much with $M$ in Figure~\ref{Mb}.  Note that the performance
of EP often deteriorates with sinusoidal sequences.

\begin{figure}[t]
    \centering
    \subfloat[\label{tpa}]{\includegraphics[width=\columnwidth]{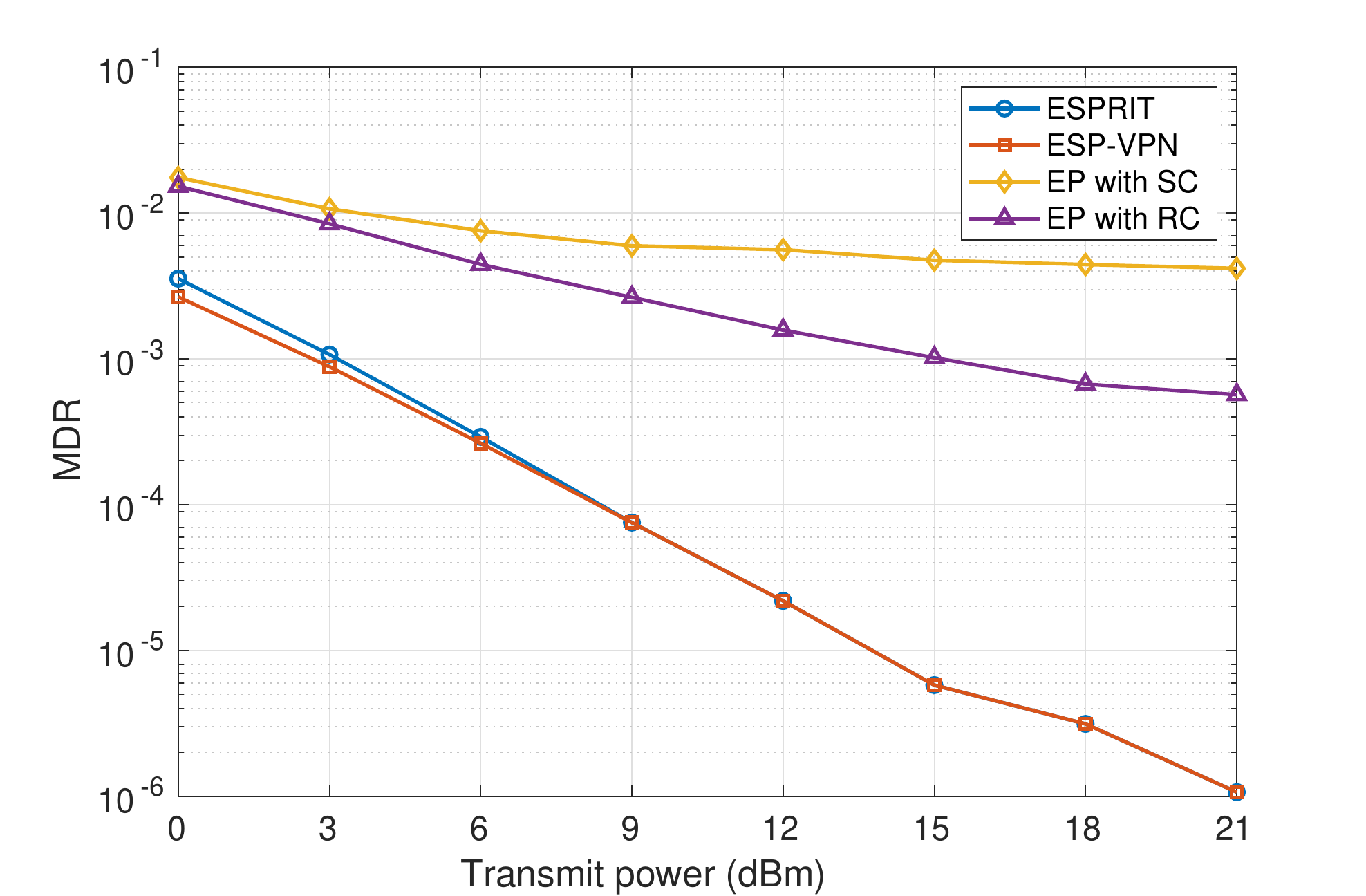}}
    
    \subfloat[\label{tpb}]{\includegraphics[width=\columnwidth]{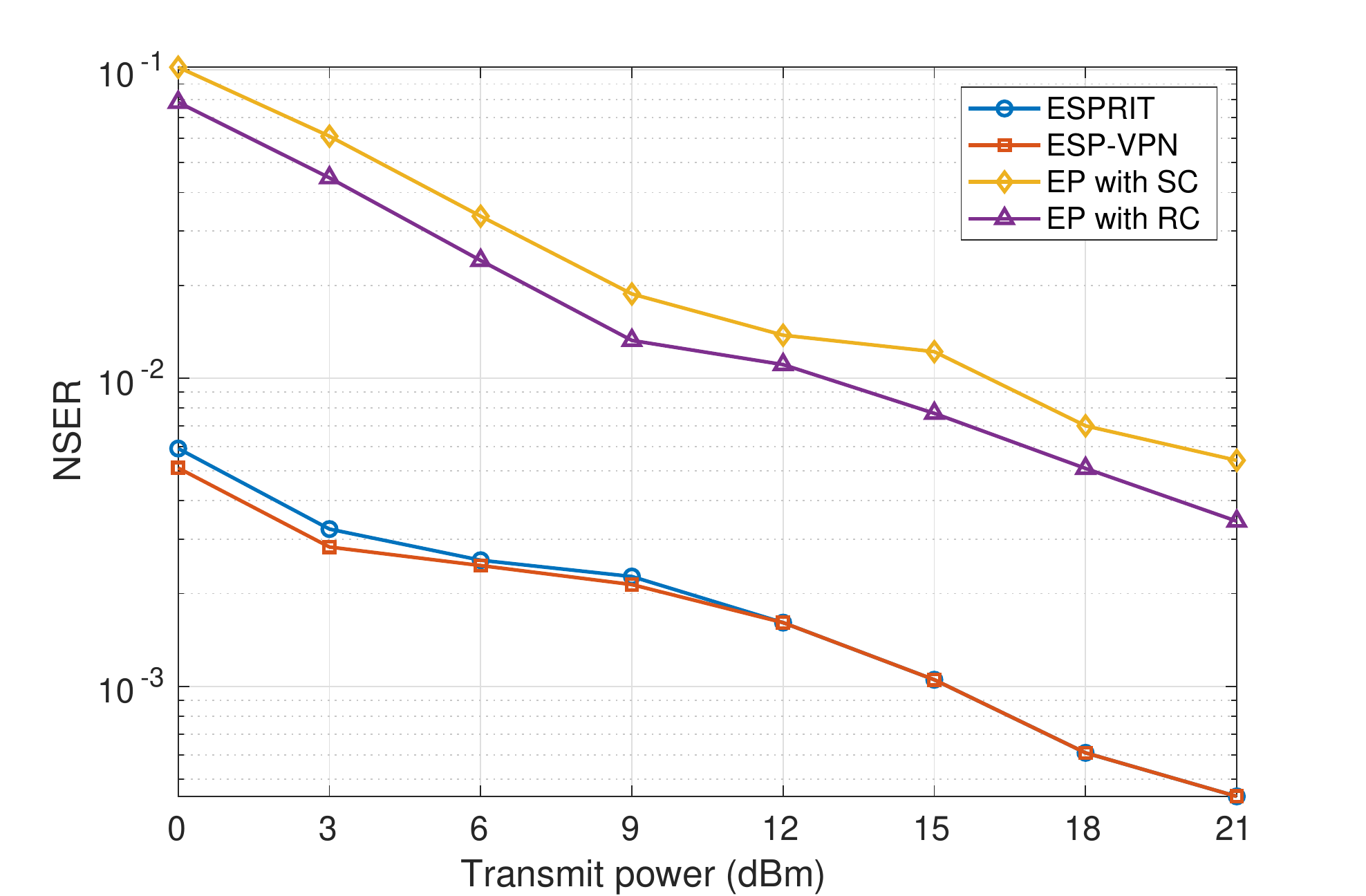}}
    
    \subfloat[\label{tpc}]{\includegraphics[width=\columnwidth]{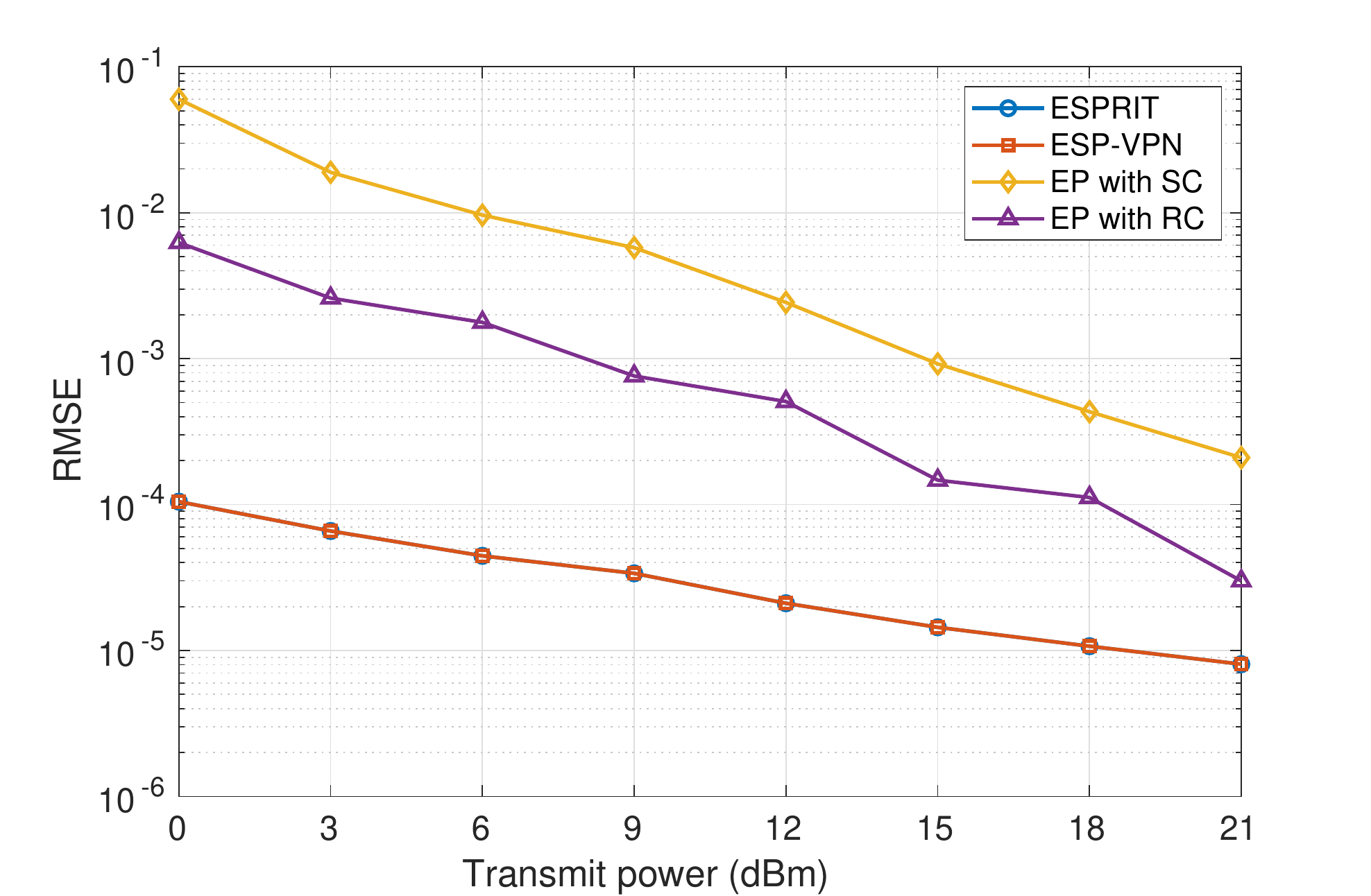}}
    \caption{(a) MDR (b) NSER (c) RMSE as functions of transmit power.}
    \label{tp}
\end{figure}

Figure~\ref{tp} plots the MDR, NSER and RMSE of different methods as functions of the transmit power, while we fix $M=64$, $p_a=0.1$. As before, the  proposed algorithm outperforms EP in all performance criteria. For transmit power  below $6$ dBm ESP-VPN provides some performance improvement over ESPRIT in terms of MDR and NSER. EP with random sequence continues to perform  better.

\begin{figure}[t]
    \centering
    \subfloat[\label{pa}]{\includegraphics[width=\columnwidth]{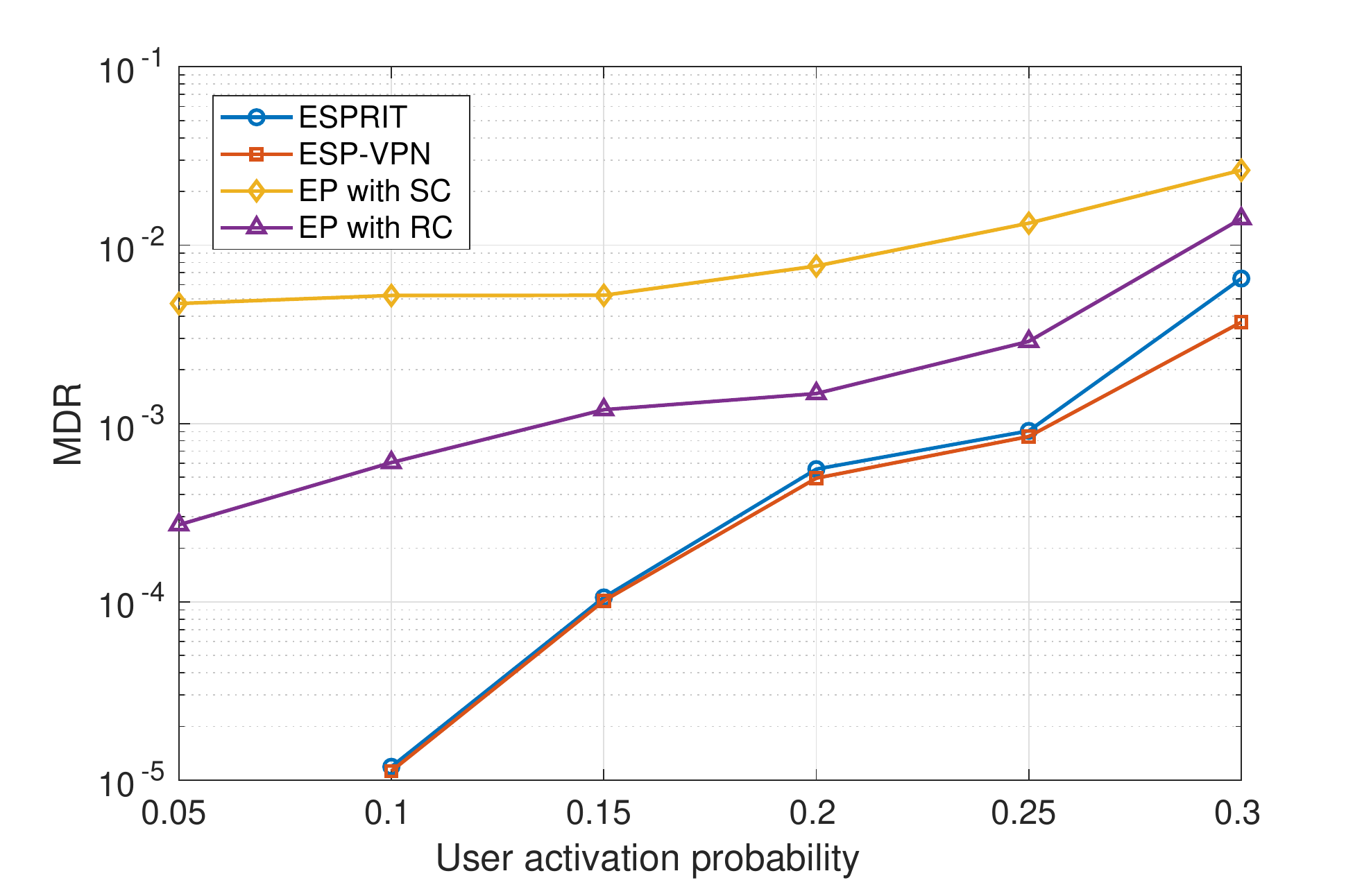}}
    
    \subfloat[\label{pb}]{\includegraphics[width=\columnwidth]{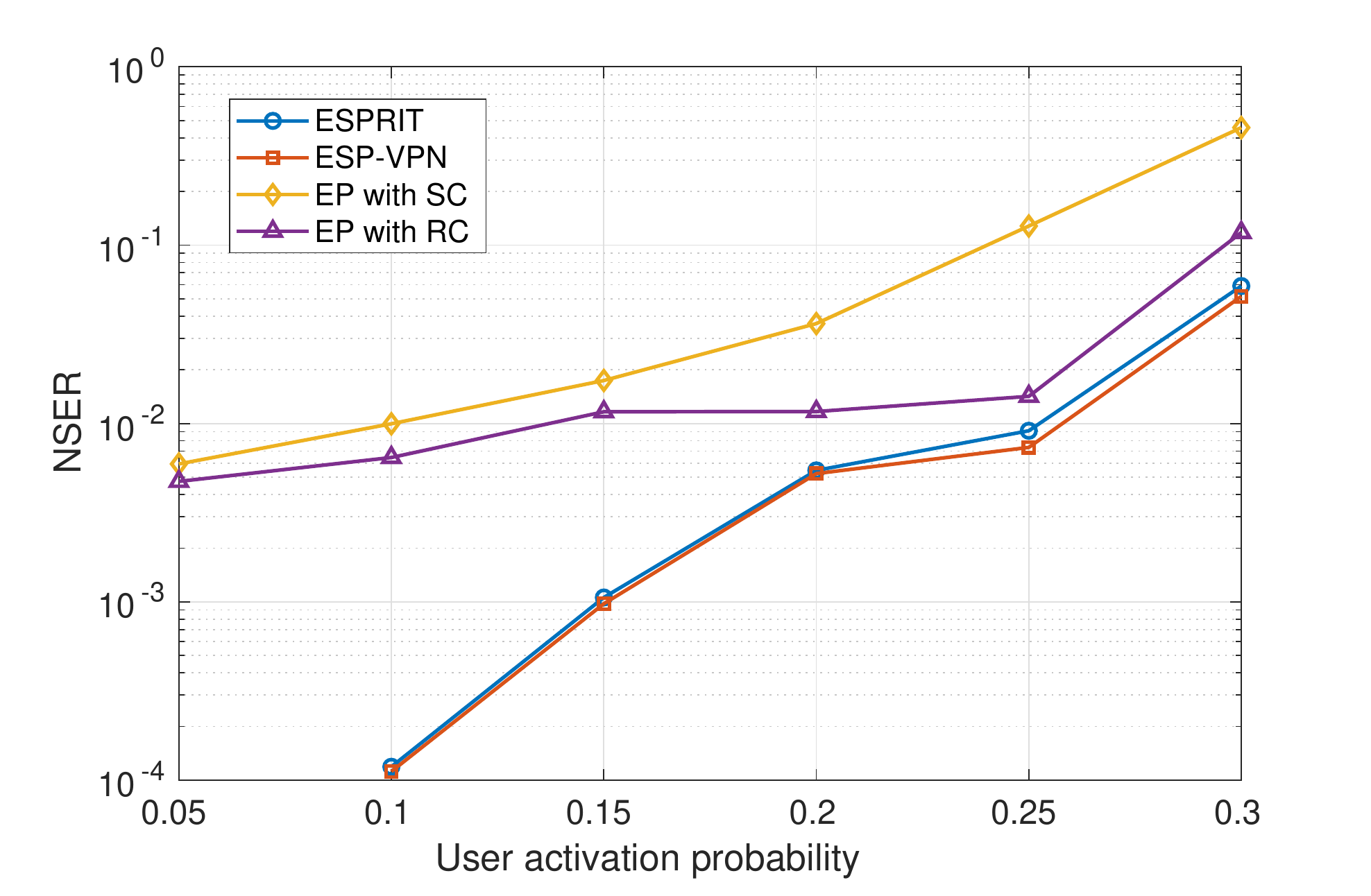}}
    
    \subfloat[\label{pc}]{\includegraphics[width=\columnwidth]{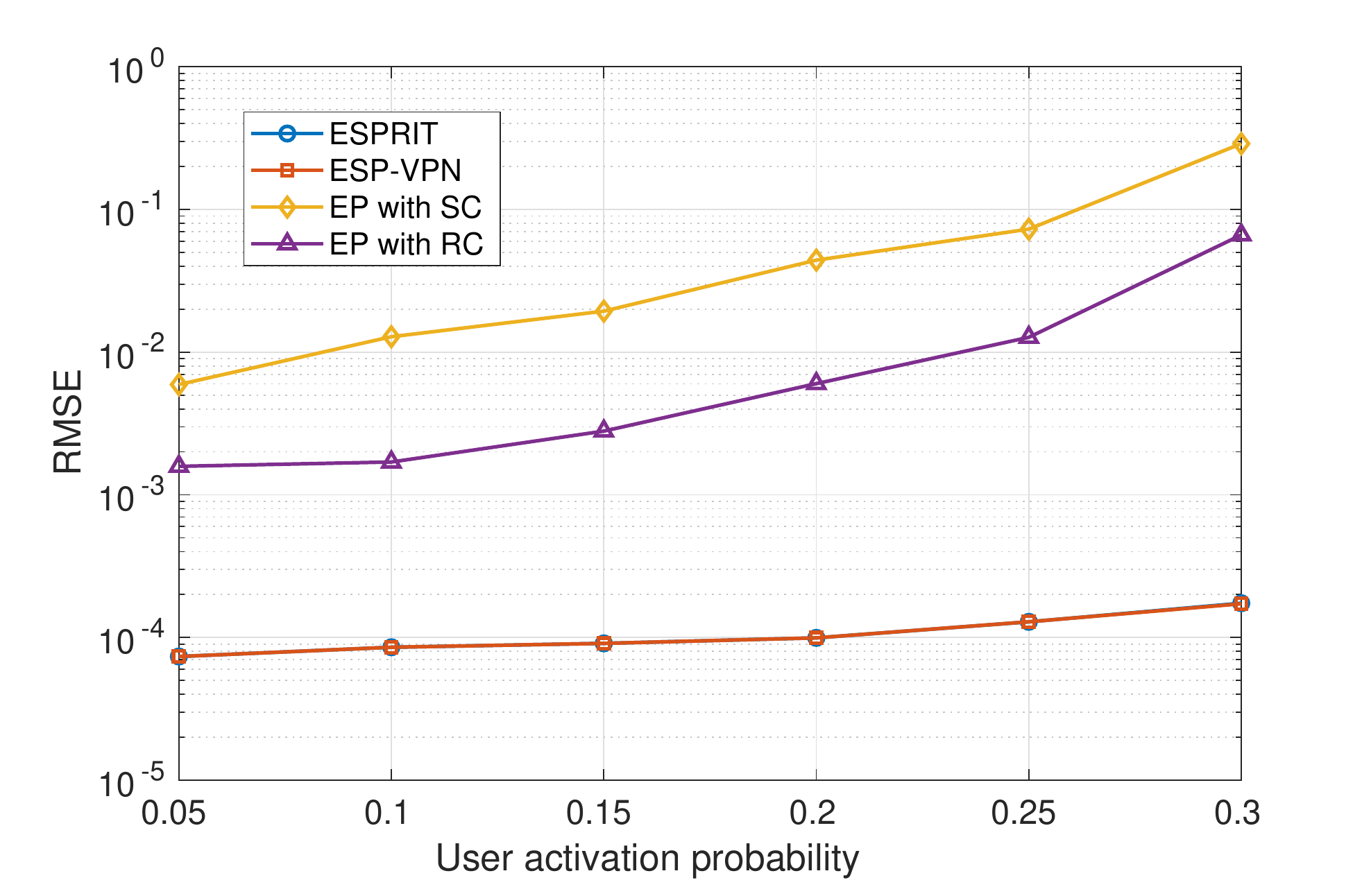}}
    \caption{(a) MDR (b) NSER (c) RMSE as functions of $p_a$.}
    \label{p}
\end{figure}

Figure \ref{p} plots the performance  metrics as functions of  user activation probability $p_a$, while the transmit power is 20 dBm and $M = 64$.  As expected, all 
methods suffer from performance degradation as $p_a$ increases  causing 
higher interference among UE. For the proposed method,  an increased number of active UEs increases probability of having active UEs with closely spaced angular frequencies. This limits the performance of ESPRIT.  In this case, ESP-VPN can 
offer slight performance improvements for  $p_a\geq 0.2$, see Figures~\ref{pa} and \ref{pb}.

\begin{figure}[t]
    \centering
    \subfloat[\label{rela}]{\includegraphics[width=\columnwidth]{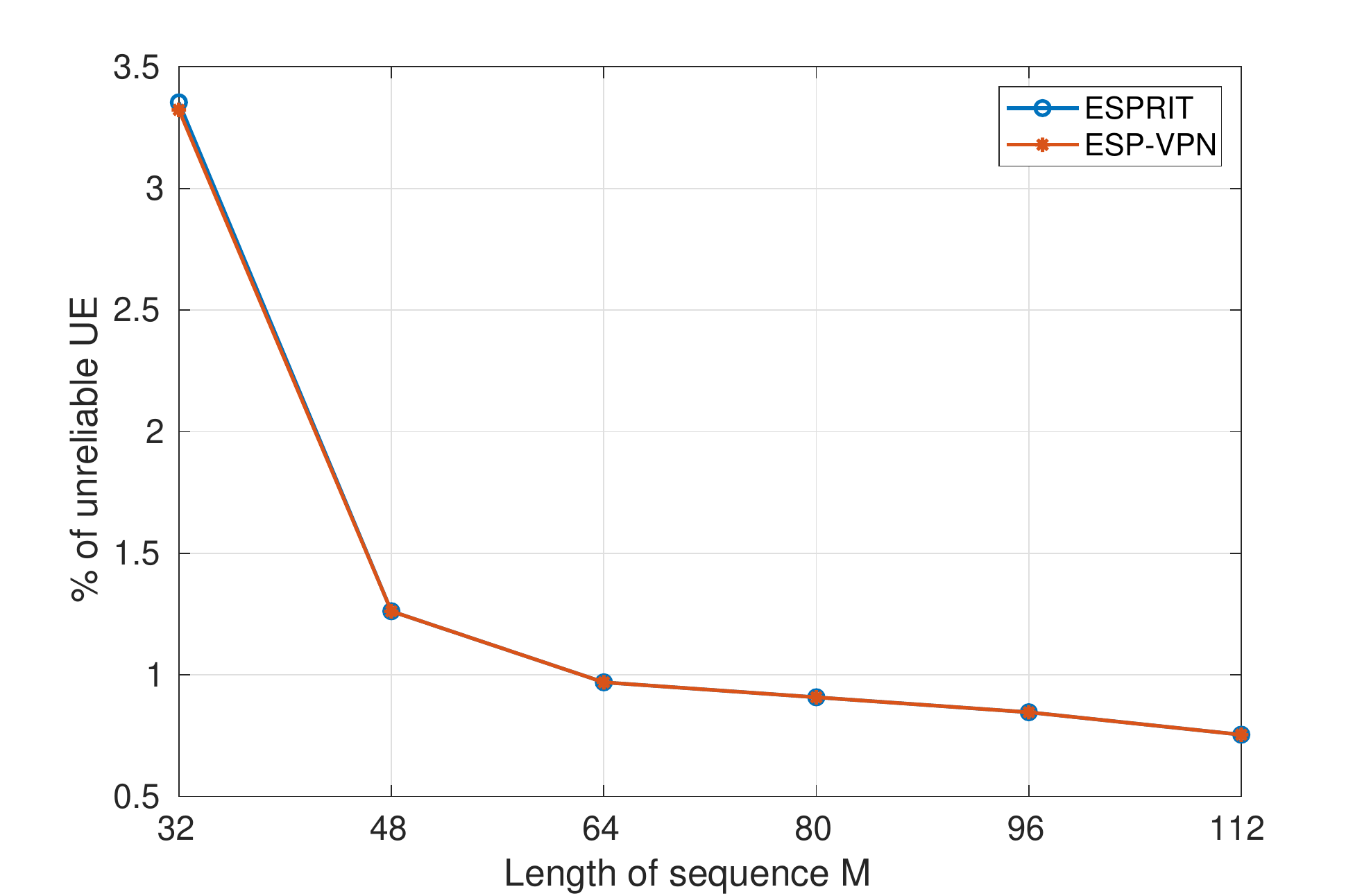}}
    
    \subfloat[\label{relb}]{\includegraphics[width=\columnwidth]{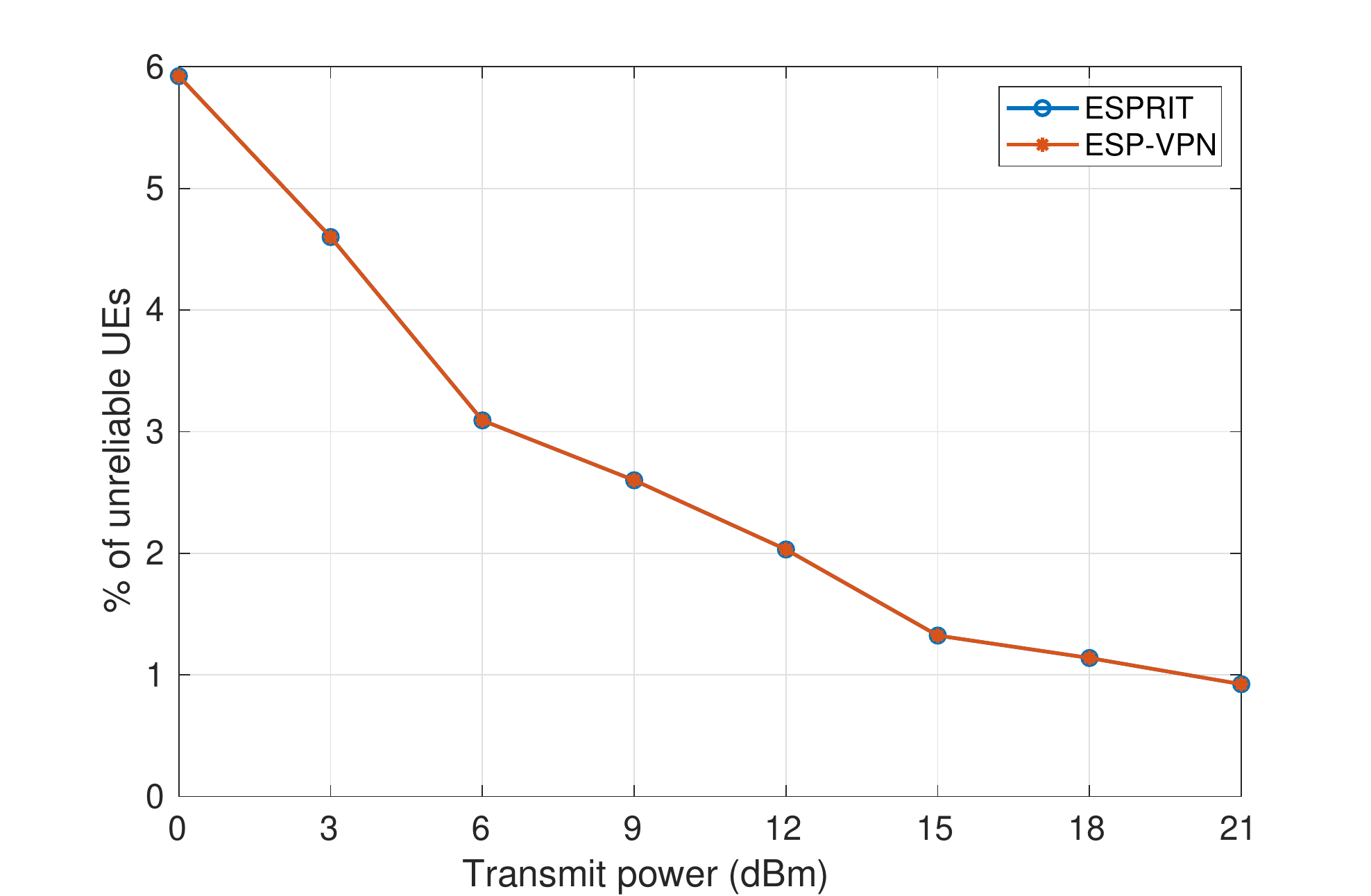}}
    
    \subfloat[\label{relc}]{\includegraphics[width=\columnwidth]{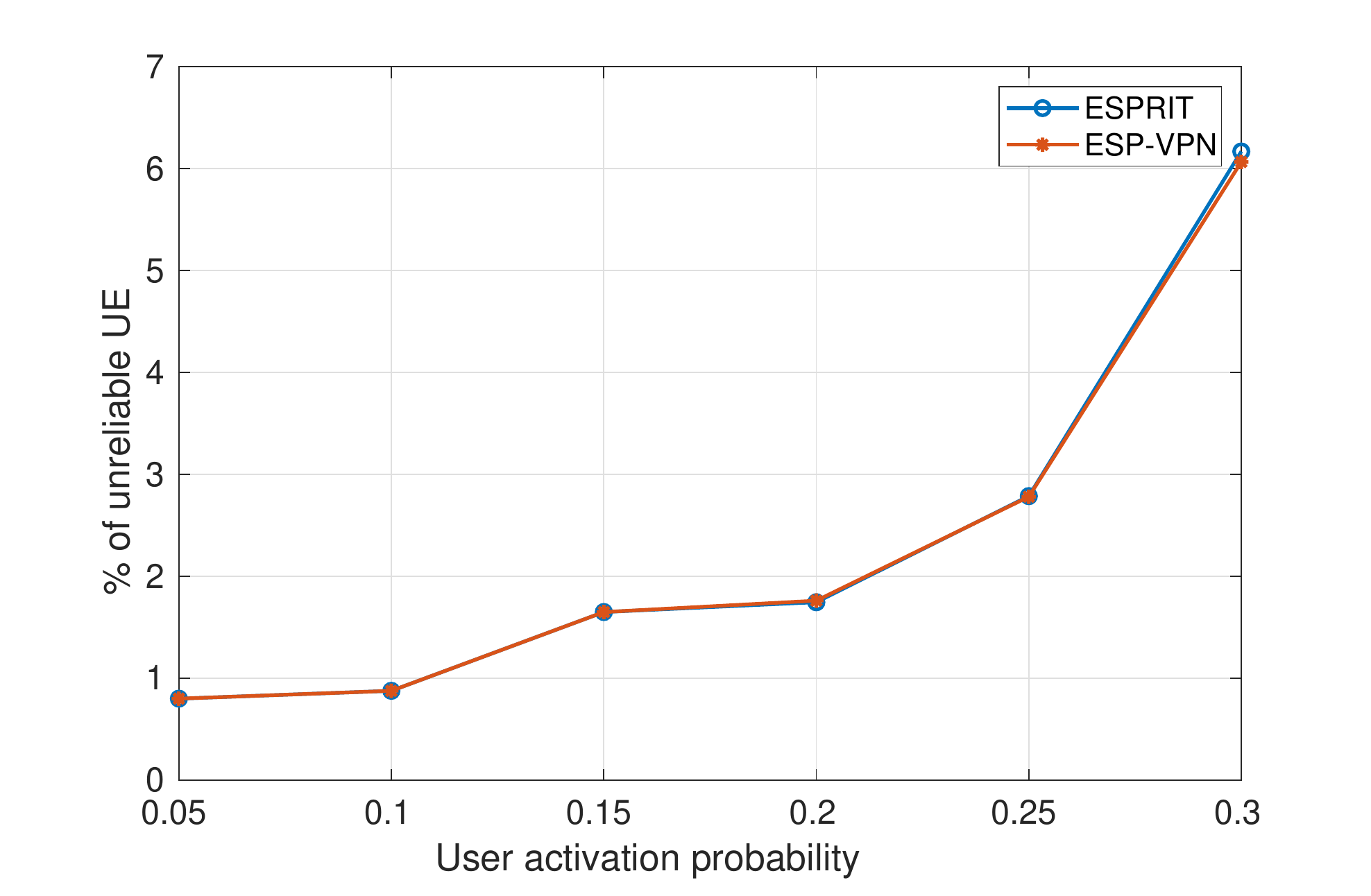}}
    \caption{\% of unreliable UEs as functions of  (a) $M$, (b) transmit power, and (c)  $p_a$.}
    \label{rel}
\end{figure}

Figure \ref{rel} shows the influence of $M, p_a,$ and transmit power on the percentage of UEs 
whose data cannot be detected reliably as per  \eqref{rell}. 
From Figure~\ref{rel}(b), where $M = 64$,  there are only 6\% of such `unreliable UEs' 
even at $0$ dBm transmit power.
 In addition, ESP-VPN offers some small improvements if $p_a$ is high.   

\begin{figure}[t]
    \centering
    \subfloat[]{\includegraphics[width=\columnwidth]{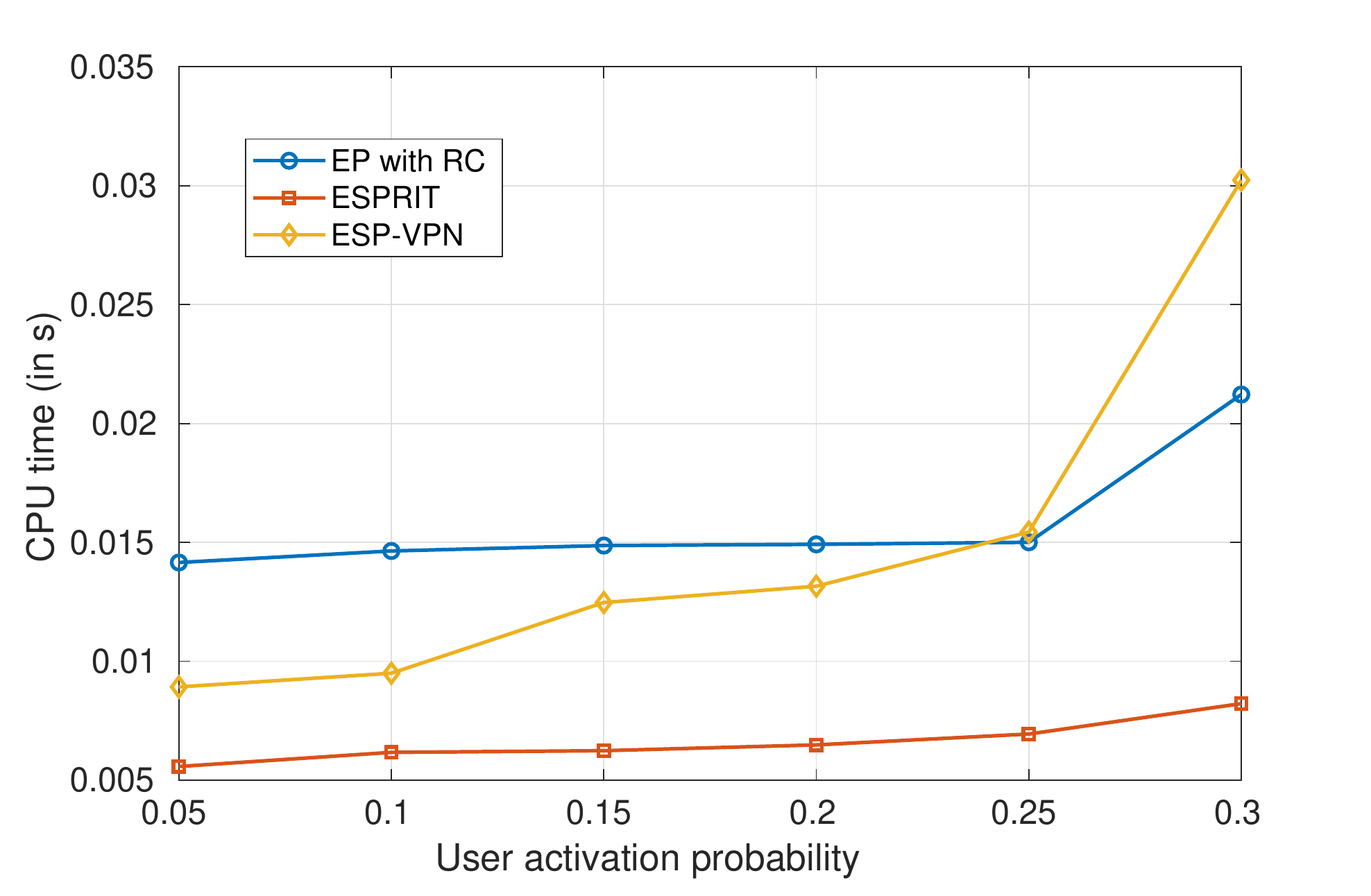}}
    
    \subfloat[]{\includegraphics[width=\columnwidth]{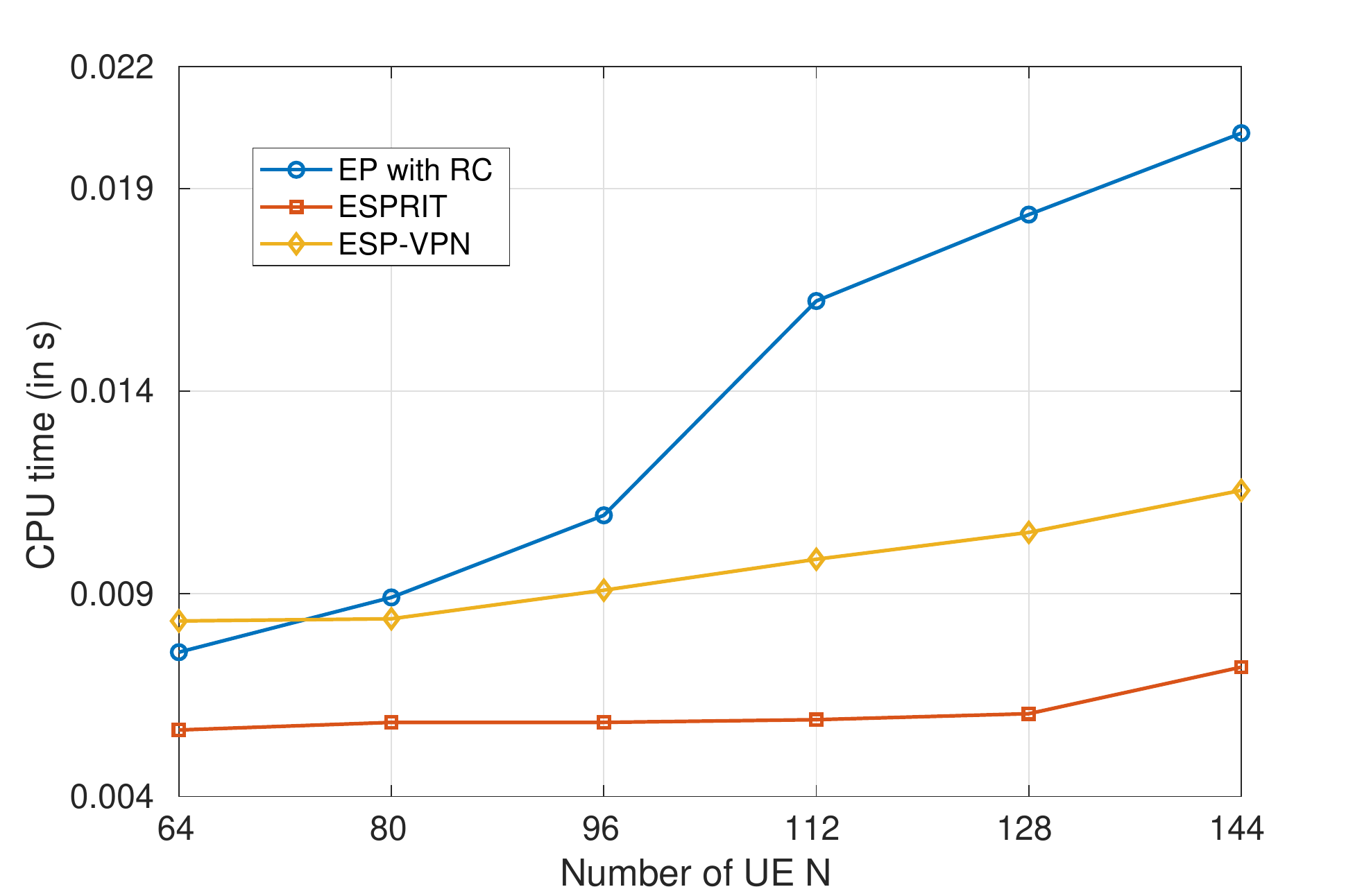}}
    \caption{Average CPU time taken  by different methods as functions of (a) $p_a$ (b) $N$.}
    \label{time}
\end{figure}

In Figure~\ref{time} we plot the average  CPU time required by different 
methods for active user identification, estimation of channel gains, and data decoding 
for $M=64$ and $20$ dBm transmit power. Here EP is run for 3 iterations (as per
\cite{ep}, the additional iterations do not improve EP's performance).
As suggested in \cite{ep}, Woodbury Identity and Cholesky decomposition is used 
for certain matrix inversion within the EP algorithm for reducing computational cost. 
All of the algorithms are implemented in MATLAB on an Intel Core i7 2600, eight-core 
computer clocked at 3.40 GHz with 16.0 GB RAM. Figure~\ref{time} demonstrates the lower 
computational requirement of proposed methods. Proposed methods (both ESPRIT and 
ESP-VPN) offer faster estimation under varying user activation probability $p_a$, although 
at $p_a\geq 0.25$, ESP-VPN shows higher complexity than EP with random sequences. 
This is because at high $p_a$, ESP-VPN in \eqref{mother1} requires higher number of iterations
 ($3\sim 4$). Note that, $p_a\geq 0.25$ is not quite practical as per \cite{empirical}, 
 where  it has been concluded that $p_a \le 0.1$ even at peak load.
 In fig. \ref{time}(b),  we fix $p_a=0.1$, and plot computation time as function of the
 total number $N$ of UEs. The computation time of EP grows quickly with $N$ at a 
 rate significantly faster than the proposed methods.

\section{Conclusions}
\label{conclusion}
In this work, we have proposed to employ sinusoidal spreading sequences for UL grant-free  mMTC system. This proposition effectively turns 
active user detection (AUD) problem into a frequency estimation problem which allows us to use non-iterative, low complexity, and accurate 
signal processing algorithms. In contrast with existing greedy and Bayesian algorithms in relevant literature, proposed method does not require 
any prior empirical assumption on channel/noise statistics and number of active users.  Extensive numerical simulations show that, in most cases, 
proposed method outperforms state-of-art EP algorithm, which  considered one of the best among the existing algorithms \cite{ep}  in terms of 
several fundamental performance metrics, that too in expense of a lower computational cost. Based on the estimated knowledge of UE activity, 
we have proposed a new method of channel estimation. This method coherently processes all data frames to minimize the adverse effects of 
measurement noise on estimation-detection performance.  This analysis also allows us to develop  a threshold aided decision rule to identify the
low SNR users, whose data symbols cannot be detected reliably.  


\bibliographystyle{IEEEtran}
\bibliography{exp}

\end{document}